\documentclass[11pt]{article}

\usepackage{amsmath}
\usepackage{amsfonts}
\usepackage{amssymb,mathrsfs}
\usepackage{theorem}
\usepackage{color}
\usepackage{graphicx}
\usepackage{caption}
\usepackage{subcaption}
\def\texitem#1{\par\smallskip\noindent\hangindent 25pt
               \hbox to 25pt {\hss #1 ~}\ignorespaces}
\numberwithin{equation}{section}

\newtheorem{theorem}{Theorem }[section]

\newenvironment{proof}[1][Proof]{\textit{#1.} }{\ \rule{0.5em}{0.5em}}
\makeatletter

\title{Worst-Case Expected Shortfall with Univariate and Bivariate Marginals}
\author{Anulekha Dhara\thanks{Engineering Systems and Design, Singapore University of Technology and Design, 8 Somapah Road, Singapore 487372.
Email: anulekha@sutd.edu.sg} \and Bikramjit Das\thanks{Engineering Systems and Design, Singapore University of Technology and Design, 8 Somapah Road, Singapore 487372. Email: bikram@sutd.edu.sg} \and Karthik Natarajan\thanks{Engineering Systems and Design, Singapore University of Technology and Design, 8 Somapah Road, Singapore 487372. Email: karthik\_natarajan@sutd.edu.sg} \footnote{This research was partly funded by the SUTD-MIT International Design Center grant number IDG31300105 on `Optimization for Complex Discrete Choice'
and the MOE Tier 2 grant number MOE2013-T2-2-168 on `Distributional Robust Optimization for
Consumer Choice in Transportation Systems'.}}
\date{}

\begin{document}
\maketitle
\begin{abstract}
Worst-case bounds on the expected shortfall risk given only limited information on the distribution of the random variables has been studied extensively in the literature. In this paper, we develop a new worst-case  bound on the expected shortfall when the univariate marginals are known exactly and additional expert information is available in terms of bivariate marginals. Such expert information allows for one to choose from among the many possible parametric families of bivariate copulas. By considering a neighborhood of distance $\rho$ around the bivariate marginals with the Kullback-Leibler divergence measure, we model the trade-off between conservatism in the worst-case risk measure and confidence in the expert information. Our bound is developed when the only information available on the bivariate marginals forms a tree structure in which case it is efficiently computable using convex optimization. For consistent marginals, as $\rho$ approaches $\infty$, the bound reduces to the comonotonic upper bound and as $\rho$ approaches $0$, the bound reduces to the worst-case bound with bivariates known exactly. We also discuss extensions to inconsistent marginals and instances where the expert information which might be captured using other parameters such as correlations.
\end{abstract}


\section{Introduction}  \label{sec1}
In recent years, there has been a growing interest in estimating worst-case bounds on the joint risk measure of a portfolio when the probability distribution of the underlying risks is uncertain. Such bounds are useful in evaluating the maximum amount of risk that a portfolio is exposed to given model uncertainty. Of particular relevance to this paper is the expected shortfall risk measure (also referred to as the Conditional Value-at-Risk (CVaR) measure) that has been proposed in the Basel Accord III as an alternative to the Value-at-Risk measure for the banking industry (see Embrechts et. al. 2014).
For a continuous random variable $\tilde{c}$ with a finite expected absolute value $\mathbb{E}_{\theta}(|\tilde{c}|) < \infty$ where $\theta$ is the probability measure and $F(\cdot)$ is the cumulative distribution function, the expected shortfall at confidence level $\alpha \in (0,1)$ is defined as:
\begin{equation*} \label{es}
\displaystyle \mbox{ES}^{\theta}_{\alpha}(\tilde{c}) = \frac{1}{1-\alpha} \int_{\alpha}^{1} \mbox{VaR}^{\theta}_{\gamma}(\tilde{c})\; \mathrm{d}\gamma,
\end{equation*}
where the Value-at-Risk at the probability level $\gamma \in (0,1)$ is defined as:
\begin{equation*} \label{var}
\displaystyle \mbox{VaR}^{\theta}_{\gamma}(\tilde{c})  = \inf\left\{c \in \mathbb{R} \ | \ F(c) \geq \gamma\right\}.
\end{equation*}
An alternative representation of the expected shortfall that was popularized by Rockafellar and Uryasev (2002) 
which is valid for any random variable $\tilde{c}$ with a finite expected absolute value is given by the optimal objective value to the following convex minimization problem:
\begin{equation*} \label{esrock}
\displaystyle \mbox{ES}^{\theta}_{\alpha}(\tilde{c}) = \inf_{\beta \in \mathbb{R}}\left(\beta + \frac{1}{1-\alpha}  \mathbb{E}_{\theta} \left[ \tilde{c} - \beta\right]^+ \right).
\end{equation*}

Consider the portfolio optimization problem:
\begin{equation} \label{eq2}
\displaystyle\min_{\mathbf{x} \in \cal{X}} \mbox{ES}^{\boldsymbol{\theta}}_{\alpha}\left(\mathbf{\tilde{c}}^T\mathbf{x} \right)
\end{equation}
where $\mathbf{x}$ is a $n$-dimensional decision vector (portfolio allocations) that is chosen in the convex set ${\cal X}$ and $\mathbf{\tilde{c}}$ is a $n$-dimensional random vector (loss of assets) with a joint distribution $\boldsymbol{\theta}$. The loss of the portfolio is given by the random variable $\mathbf{\tilde{c}}^T\mathbf{x} = \sum_{i} \tilde{c}_ix_i$ where the expected shortfall captures the risk of the portfolio. The expected shortfall risk measure has several attractive mathematical properties. Firstly, the portfolio optimization problem with an expected shortfall risk measure is a convex optimization problem (unlike the VaR measure) and formulated as follows:
\begin{equation} \label{eq2aa}
\displaystyle\min_{\mathbf{x} \in \cal{X},\beta \in \mathbb{R}} \left(\beta+\frac{1}{1-\alpha}\mathbb{E}_{\boldsymbol{\theta}} \left[\mathbf{\tilde{c}}^T\mathbf{x} - \beta\right]^+ \right).
\end{equation}
Moreover, expected shortfall is a coherent risk measure (see Artzner et. al. 1999) 
and encourages risk diversification. However, there remains challenges in using this risk measure in portfolio optimization. Lim et. al. (2011) 
have shown that the estimation error with the expected shortfall risk measure tends to be magnified since the model is very sensitive to the tail of the return distribution. Hanasusanto et. al. (2016) 
have shown that computing the expected value of the non-negative part of a linear combination of random variables of the form
\begin{equation*}
\mathbb{E}_{\boldsymbol{\theta}} \left[\mathbf{\tilde{c}}^T\mathbf{x} - \beta\right]^+,
\end{equation*}
for fixed values of $\mathbf{x}$ and $\beta$ is \#P-hard, even for uniformly distributed independent random variables. 

One approach to tackle these difficulties is to allow for the joint probability distribution $\boldsymbol{\theta}$ to lie in a possible set of distributions $\boldsymbol{\Theta}$. Given this set of distributions, the {distributionally robust portfolio optimization problem} is formulated as
\begin{equation} \label{eq3}
\displaystyle\min_{\mathbf{x} \in \cal{X}}\: \max_{\boldsymbol{\theta} \in \boldsymbol{\Theta}}\: \mbox{ES}^{\boldsymbol{\theta}}_{\alpha}\left(\mathbf{\tilde{c}}^T\mathbf{x} \right),
\end{equation}
where the portfolio allocation vector is chosen to minimize the worst-case expected shortfall risk over all possible distributions in the set $\boldsymbol{\Theta}$. Under specific assumptions on the set $\boldsymbol{\Theta}$, formulation (\ref{eq3}) has been shown to be efficiently solvable. For discrete distributions, examples of the set $\boldsymbol{\Theta}$ for which the worst-case expected shortfall and the corresponding distributionally robust optimization problem are efficiently solvable are:
\texitem{(a)} Sets of distributions with fixed univariate marginals (see R\"{u}schendorf 1983), 
\texitem{(b)} Sets of distributions with fixed nonoverlapping multivariate marginals (see Doan and Natarajan 2012, 
Embrechts and Pucetti 2006), 
\texitem{(c)} Sets of distributions with fixed overlapping multivariate marginals where the structure of the overlapping marginals satisfies a regularity property (see Doan, Li and Natarajan 2015), 
\texitem{(d)} Sets of distributions where the probabilities of the scenarios are assumed to lie in a box uncertainty set or an ellipsoidal uncertainty set (see Zhu and Fukushima 2009), 
\texitem{(e)} Sets of distributions where the probabilities of the scenarios are assumed to lie in a uncertainty set defined using the $\phi$-divergence measure around a nominal probability distribution (see Ben-Tal et. al. 2013). 
This includes Kullback-Leibler divergence, Burg entropy and $\chi^2$-distance as special cases.  

\noindent Note that in cases (b), (c), (d) and (e), the representation of the set $\boldsymbol{\Theta}$ might itself be exponential in the number of random variables. For example, if each of the $n$ random variables takes $m$ possible values, the joint distribution can take up to $m^n$ possible values. Hence, efficient solvability in cases (b)-(e) typically refers to an algorithm that solves the problem in polynomial time in the number of support points of the joint discrete distribution.

\subsection{Motivation}
In the literature, several approaches have been proposed to model distributions where information on the marginals is available and only limited dependence information is known.
One popular approach is to find the maximum entropy distribution (see Jaynes 1957) 
where the joint distribution is chosen to maximize the Shannon entropy measure (see Shannon 1948) 
satisfying the given information. For example, given univariate marginals, the maximum entropy distribution is the independent distribution. Similarly, given bivariate information which forms a tree structure, the maximum entropy distribution is a conditionally independent distribution on a tree, referred to as a Markov (Chow-Liu 1968) 
tree distribution. An alternate approach is to use the theory of copulas in which the joint distribution is modeled in terms of the univariate marginals and a copula (see Sklar 1959). 
Specifying and estimating the copula in high dimensions however still remains challenging. A closely related model for distributions in high dimensions that builds on bivariate copulas is a vine copula (see Bedford and Cooke 2002). 
Unlike Chow-Liu tree distributions on $n$ random variables where only dependence on $n-1$ edges can be specified, in vine copulas it is possible to specify additional dependence information among the variables using bivariate copula. The original motivation for using vine copula stemmed from the need to incorporate expert opinions on dependencies into the specification of the distribution. In our work, instead of describing the joint distribution that is consistent with univariate and bivariate information by specifying a copula structure, we focus on finding the worst-case joint distribution for the expected shortfall risk measure. 

In this paper, we assume that the univariate marginal distributions are precisely known but allow for the information on the bivariate distributions to be generated in a variety of ways including historical data which might involve noisy, imprecise or incomplete data observations and expert opinions. Hence it is quite possible that there  exist no distribution which is consistent with the given univariate and bivariate marginal information. For example in portfolio optimization, correlation between the stocks is often used to model dependencies. These correlations may be measured between pairs of stocks over different periods of time. Furthermore, experts often provide estimates on correlations between pairs of stocks based on their forecasts. Combining different expert views can lead to a negative semidefinite or indefinite correlation matrix. This has led to an interest in the ``closest correlation matrix problem" where the objective is to find a positive semidefinite correlation matrix that is closest to the  given matrix. While much work has been done towards finding and constructing  the closest correlation matrix (see Higham 2002, 
Qi and Sun 2010), 
far fewer work involve studying its effect on portfolio optimization.

Our approach is closely related to four streams of work. Glasserman and Yang (2016) 
recently proposed an optimization formulation to compute the worst-case credit valuation adjustment risk by combining models of market and credit risk. In their approach, the marginal distributions are held fixed while the worst-case risk is computed by penalizing for deviation with respect to a bivariate reference model of the dependence between the market and credit risk models. While their optimization formulations are developed for $n = 2$, we generalize this to arbitrary $n$ while preserving polynomial time solvability. The second stream of work is the model proposed by Ben-Tal et. al (2013) 
who studied robust optimization problems where the set of joint distributions lies in a neighborhood of a  reference probability distribution with distance that is defined using the notion of $\phi$-divergence. Their convex optimization formulation is polynomial in the number of support points of the joint distribution but can be  exponential in the number of random variables. On the other hand in our work by considering neighborhoods of lower dimensional probability distributions, specifically bivariate marginals and assuming a tree structure, we develop a convex reformulation that grows polynomially in the number of random variables. The third stream of results is by Roughgarden and Kearns (2013) 
who discuss the hardness of verifying if a joint distribution exists with the given univariate and bivariate marginals and the corresponding problem of finding the closest consistent marginals to inconsistent marginals. While their focus is on showing the NP-hardness of these problems on a complete graph, we show that these problems are easy to solve on a tree and furthermore it is possible to evaluate robust bounds. Finally, our results builds on the earlier work of Doan, Li and Natarajan (2015) 
who develop bounds on expected shortfall using univariate and bivariate marginal information. In contrast to their work, we allow for the marginals to be inconsistent and even when consistent, we provide flexibility in modeling the confidence in the expert opinions by considering a neighborhood around the marginals. Thus, we explicitly model the trade-off between conservatism in the worst-case risk measure and confidence in the expert information.



The outline of the paper is as follows. In Section 2, we provide a formal description of the set of distributions $\boldsymbol{\Theta}$ with given univariate marginals $\boldsymbol{\mu}_i$ and expert information in terms of bivariate marginals $\boldsymbol{\mu}_{ij}$. We study the problem of finding the closest consistent marginals by finding the smallest $\rho$-neighborhood using the Kullback-Leibler divergence around the given bivariate marginals such that $\boldsymbol{\Theta}$ is nonempty. We also provide a polynomial-sized convex reformulation for this problem by  assuming a tree structure. We also discuss several extensions in this section. In Section 3, we define the uncertainty set of distributions using a $\rho$-neighborhood of the bivariate distributions such that $\boldsymbol{\Theta}$ is nonempty. We provide a convex optimization problem to find the worst-case upper bound on the expected shortfall. We show that the bound is efficiently computable when the bivariate marginal information forms a tree structure. In Section 4, we present numerical experiments. All proofs are provided in the Appendix.

\section{Consistent Marginals in $\boldsymbol{\Theta}$} \label{sec2}

In this section, we provide a formal description of the set of distributions $\boldsymbol{\Theta}$. We restrict our attention to univariate and bivariate marginal information.  Denote the random vector by $\mathbf{\tilde{c}} = (\tilde{c}_1, \tilde{c}_2, \ldots, \tilde{c}_n)$ and define the index set  $\mathcal{N} = \{1,2,\ldots,n\}$. Let $\mathcal{C}_i$ denote the set of values taken by $\tilde{c}_i$ for $i \in \mathcal{N}$. For $i \in \mathcal{N}$, let $\boldsymbol{\mu}_i$ denote the given univariate distribution where $\mu_{i}(c_i) = \mathbb{P}_{\boldsymbol{\theta}}(\tilde{c}_i = c_i)$ for $c_i \in \mathcal{C}_i$. Let $\mathcal{N}' \subseteq \mathcal{N} \times \mathcal{N}$ denote the index set for the bivariate marginal information. Then for $(i,j) \in \mathcal{N}'$ with $i <j$, let $\boldsymbol{\mu}_{ij}$ denote the bivariate marginal where $\mu_{ij}(\mathbf{c}_{ij}) = \mathbb{P}_{\boldsymbol{\theta}}(\mathbf{\tilde{c}}_{ij} = \mathbf{c}_{ij}) = \mathbb{P}_{\boldsymbol{\theta}}(\tilde{c}_i = c_i, \tilde{c}_j = c_j)$ with $\mathbf{\tilde{c}}_{ij} = (\tilde{c}_i,\tilde{c}_j)$ and $\mathbf{c}_{ij} = (c_i,c_j)$ for $c_i \in \mathcal{C}_i$ and $c_j \in \mathcal{C}_j$. We assume that the bivariate marginal information is obtained from expert information. Such bivariate marginals might be chosen using one of many possible parametric families of bivariate copula. We also discuss extensions when such expert information on bivariates might be captured using simpler parameters such as correlations.  
The associated graph with this set of marginals consist of the vertex set $\mathcal{N}$ where each vertex $i$ corresponds to the random variable $\tilde{c}_i$ and the arc set $\mathcal{N}'$ where each undirected edge $(i,j)$ between vertex $i$ and $j$ corresponds to a pair of random variables $\tilde{c}_i$ and $\tilde{c}_j$ for which the bivariate marginal is provided. Let $(\mathcal{N},\mathcal{N}')$ denote the corresponding graph.

Let $\boldsymbol{\Theta}$ denote the Fr\'{e}chet class of joint distributions with the given univariate and bivariate marginals defined as
\begin{equation*}\label{feasible}
\boldsymbol{\Theta} = \{\boldsymbol{\theta}: \textrm{proj}_i(\boldsymbol{\theta}) = \boldsymbol{\mu}_i, \forall i \in \mathcal{N}, \textrm{proj}_{ij}(\boldsymbol{\theta}) = \boldsymbol{\mu}_{ij},\forall (i,j) \in \mathcal{N}'\},
\end{equation*}
where $\textrm{proj}_i(\cdot)$ and $\textrm{proj}_{ij}(\cdot)$ denote the projection of the joint distribution to the $i$-th and $(i,j)$-th variables. An important question is to identify conditions on the univariate and bivariate marginals which would guarantee that the set $\boldsymbol{\Theta}$ is nonempty. A necessary set of conditions to guarantee the set is nonempty is the consistency of the univariate and bivariate marginals which is given by the equality conditions:
\begin{equation} \label{feasible00}
\begin{array}{rlll}
\displaystyle\sum_{c_j \in \mathcal{C}_j} {\mu}_{ij}(\mathbf{c}_{ij}) & = &  {\mu}_{i}(c_i), &  \forall c_i \in \mathcal{C}_i, \forall i \in \mathcal{N}: (i,j) \in \mathcal{N}', \\
\displaystyle\sum_{c_i \in \mathcal{C}_i} {\mu}_{ij}(\mathbf{c}_{ij}) & = & {\mu}_{j}(c_j), & \forall c_j \in \mathcal{C}_j, \forall j \in \mathcal{N}: (i,j) \in \mathcal{N}'.
\end{array}
\end{equation}
However, the consistency of univariate and bivariate marginals is not a sufficient condition to guarantee the nonemptiness of $\boldsymbol{\Theta}$ (see Vorob'ev 1962). 
For example, consider $\mathcal{N} = \{1,2,3\}$ with binary random variables. The univariate marginals are given as ${\mu}_{1}(0) = {\mu}_{1}(1) = 1/2$, ${\mu}_{2}(0) = {\mu}_{2}(1) = 1/2$, ${\mu}_{3}(0) = {\mu}_{3}(1) = 1/2$ and the bivariate marginals are given as ${\mu}_{12}(0,1) = {\mu}_{12}(1,0) = 1/2$, ${\mu}_{23}(0,1) = {\mu}_{23}(1,0) = 1/2$, ${\mu}_{13}(0,1) = {\mu}_{13}(1,0) = 1/2$. Though the univariate and bivariate marginals are consistent in this example, the set $\boldsymbol{\Theta}$ is empty. The nonemptiness of the Fr\'{e}chet class of distributions $\boldsymbol{\Theta}$ is ensured under the consistency of multivariate marginals when the graph satisfies a graph theoretic condition known as the ``running intersection property''. One such graph structure is a tree. The consistency of the univariate and bivariate marginals in this case will ensure that $\boldsymbol{\Theta}$ is nonempty. The tree structure plays an important role in the present work as it simplifies the formulations as discussed next.

\subsection{Closest Consistent Marginals}
In this section, we consider the problem of finding the closest consistent marginals when the original set of distribution $\boldsymbol{\Theta}$ is empty. To do so, we find the minimal  perturbation $\rho$ for all the given bivariate marginals such that the new bivariate marginals $\{\boldsymbol{\theta}_{ij}\}_{(i,j) \in \mathcal{N}'}$ lie within a $\rho$-neighborhood of $\{\boldsymbol{\mu}_{ij}\}_{(i,j) \in \mathcal{N}'}$ and there exists a joint distribution consistent with the given univariate marginals $\{\boldsymbol{\mu}_i\}_{i \in \mathcal{N}}$ and the new bivariate marginals $\{\boldsymbol{\theta}_{ij}\}_{(i,j) \in \mathcal{N}'}$. To define the $\rho$-neighborhood, we make use of the notion of Kullback-Leibler (KL)-divergence. For any two non-negative vectors $\mathbf{p} = (p_1, \ldots, p_m)^T$ and $\mathbf{q} = (q_1, \ldots, q_m)^T$, the KL-divergence is defined by: 
\begin{center}
$I_{KL}(\mathbf{p},\mathbf{q}) = \displaystyle\sum_{i=1}^m p_i\; \log\left(\frac{p_i}{q_i}\right).$
\end{center}
 The KL-divergence is an example of $\phi$-divergence (see Pardo 2006) 
 which has recently gained popularity in the area of distributionally robust optimization (see Ben-Tal et al. 2013) 
 where the set of probability distributions is defined to be the $\rho$-neighborhood of a reference probability distribution.

For verifying the nonemptiness of $\boldsymbol{\Theta}$, we need to check for the existence of a joint distribution $\boldsymbol{\theta}$ with the given univariate marginals $\{\boldsymbol{\mu}_i\}_{i \in \mathcal{N}}$ and bivariate marginals $\{\boldsymbol{\mu}_{ij}\}_{(i,j) \in \mathcal{N}'}$. If such a $\boldsymbol{\theta}$ exists, the minimum deviation is given by $\rho = 0$ else $\rho > 0$. 
We next formulate an optimization problem to find the smallest neighborhood of the given bivariate marginals which guarantees the nonemptiness of $\boldsymbol\Theta$. Let $\boldsymbol{\mathcal{C}} = {\cal C}_1 \times {\cal C}_2 \times \ldots {\cal C}_n$ be the set of all possible  realizations of $\mathbf{\tilde{c}}$ and $\boldsymbol\theta(\mathbf{c}) = \mathbb{P}(\mathbf{\tilde{c}} = \mathbf{c})$ for $\mathbf{c} \in \boldsymbol{\mathcal{C}}$. The problem of finding the closest consistent marginals using the KL-divergence measure is formulated as the solution to a convex optimization problem:
\begin{equation} \label{over0}
\begin{array}{rll}
\displaystyle\min_{\rho,\;\boldsymbol\theta,\;\boldsymbol {\theta}_{ij}} & \rho & \\
\mbox{subject to} &\displaystyle\sum_{\mathbf{c} \in \boldsymbol{\mathcal{C}}|c_i} \boldsymbol{\theta}(\mathbf{c}) = {\mu}_{i}(c_i), & \forall i \in \mathcal{N}, \forall c_i \in \mathcal{C}_i,  \\
&\displaystyle\sum_{\mathbf{c} \in \boldsymbol{\mathcal{C}}| \mathbf{c}_{ij}} \boldsymbol{\theta}(\mathbf{c}) = {\theta}_{ij}(\mathbf{c}_{ij}), &\forall (i,j) \in \mathcal{N}',\forall \mathbf{c}_{ij} \in \mathcal{C}_i \times \mathcal{C}_j,  \\
& \displaystyle\sum_{\mathbf{c} \in \boldsymbol{\mathcal{C}}} \boldsymbol{\theta}(\mathbf{c}) = 1, & \\
 & \displaystyle \boldsymbol{\theta}(\mathbf{c}) \geq 0, & \forall \mathbf{c} \in \boldsymbol{\mathcal{C}}, \\
&I_{KL}\left(\boldsymbol{\theta}_{ij},\boldsymbol{\mu}_{ij}\right) \leq \rho, & \forall (i,j) \in \mathcal{N}'.\\
\end{array}
\end{equation}
In formulation (\ref{over0}), the objective is to minimize the distance of the neigborhood given by $\rho$. The first four constraints ensure that there exists a joint distribution with univariate $\boldsymbol{\mu}_i$ and bivariate $\boldsymbol{\theta}_{ij}$ while the last constraint ensures that the bivariate $\boldsymbol{\theta}_{ij}$ lies in a $\rho$-neighborhood of $\boldsymbol{\mu}_{ij}$. Note that for consistent marginals $\boldsymbol{\mu}_i$ and $\boldsymbol{\mu}_{ij}$ such that a joint distribution $\boldsymbol{\theta}$ exists, the optimal solution of (\ref{over0}) is $\rho = 0$. 

The number of variables in the convex optimization problem in  (\ref{over0}) is however exponential in the number of variables $n$. We next identify an instance where the closest consistency problem can be solved as a polynomial sized convex optimization problem. The result is provided next and the proof is provided in the Appendix.

\begin{theorem} \label{thm2.1}
Consider the given univariate and bivariate marginals $\{\boldsymbol{\mu}_i\}_{i \in \mathcal{N}}$ and $\{\boldsymbol{\mu}_{ij}\}_{(i,j) \in \mathcal{N}'}$  where the graph associated with $(\mathcal{N},\mathcal{N}')$ is a tree. Then the solution to the following convex optimization problem finds the minimal $\rho$-neighborhood for which there exists a joint distribution:

\begin{equation} \label{over1}
\begin{array}{rll}
\displaystyle\min_{\rho,\;\boldsymbol{\theta}_{ij}} & \rho \\
\textrm{subject to} & \displaystyle\sum_{c_j \in \mathcal{C}_j} \theta_{ij}(\mathbf{c}_{ij}) = \mu_i(c_i), & \forall i \in \mathcal{N}: (i,j) \in \mathcal{N}',\forall c_i \in \mathcal{C}_i, \\
& \displaystyle\sum_{c_i \in \mathcal{C}_i} \theta_{ij}(\mathbf{c}_{ij}) = \mu_j(c_j), & \forall j \in \mathcal{N}: (i,j) \in \mathcal{N}',\forall c_j \in \mathcal{C}_j, \\
 & \displaystyle \sum_{\mathbf{c}_{ij} \in \mathcal{C}_i \times \mathcal{C}_j} {\theta}_{ij}(\mathbf{c}_{ij}) = 1, & \forall (i,j) \in \mathcal{N}',   \\ 
 & \displaystyle {\theta}_{ij}(\mathbf{c}_{ij}) \geq 0, & \forall (i,j) \in \mathcal{N}', \forall \mathbf{c}_{ij}\in {\cal C}_i \times {\cal C}_j,     \\
&\displaystyle I_{KL}(\boldsymbol{\theta}_{ij},\boldsymbol{\mu}_{ij}) \leq \rho, &  \forall (i,j) \in \mathcal{N}'.
\end{array}
\end{equation}
\end{theorem}
In formulation (\ref{over0}), the number of decision variables is $O(m^n)$ where $m$ is the number of possible values that each of the $n$ random variables takes. On the other hand in formulation (\ref{over1}), the number of decision variables is only $O(nm^2)$.

\subsection{Extensions}
In this section, we relate Theorem \ref{thm2.1} to three existing models and discuss possible extensions. 

Roughgarden and Kearns (2013) 
considered the problem of finding a minimal $\rho$-neighborhood over which a consistent joint distribution $\boldsymbol{\theta}$ exists which is obtained by perturbing both the univariate marginals $\{\boldsymbol{\mu}_i\}_{i \in \mathcal{N}}$ and the bivariate marginals $\{\boldsymbol{\mu}_{ij}\}_{(i,j) \in \mathcal{N}'}$. In their work, the authors formulated a linear program by minimizing the $L_1$-norm distance instead of the KL-divergence measure that we use in (\ref{over0}). However since they consider a general graph structure for the bivariates, their problem is NP-hard to solve. We can build on their approach in cases when the univariate marginal information is not reliable by also perturbing the univariate marginals using the KL-divergence measure. The convex optimization problem (\ref{over1}) for the tree structure in this case can be extended to the following formulation:
\begin{equation} \label{over1n}
\begin{array}{rll}
\displaystyle\min_{\rho,\;\boldsymbol{\theta}_{i},\;\boldsymbol{\theta}_{ij}} & \rho \\
\textrm{subject to} & \displaystyle\sum_{c_j \in \mathcal{C}_j} \theta_{ij}(\mathbf{c}_{ij}) = \theta_i(c_i), & \forall i \in \mathcal{N}: (i,j) \in \mathcal{N}',\forall c_i \in \mathcal{C}_i, \\
& \displaystyle\sum_{c_i \in \mathcal{C}_i} \theta_{ij}(\mathbf{c}_{ij}) = \theta_j(c_j), & \forall j \in \mathcal{N}: (i,j) \in \mathcal{N}',\forall c_j \in \mathcal{C}_j, \\
 & \displaystyle \sum_{\mathbf{c}_{ij} \in \mathcal{C}_i \times \mathcal{C}_j} {\theta}_{ij}(\mathbf{c}_{ij}) = 1, & \forall (i,j) \in \mathcal{N}',   \\ 
 & \displaystyle {\theta}_{ij}(\mathbf{c}_{ij}) \geq 0, &  \forall (i,j) \in \mathcal{N}', \forall \mathbf{c}_{ij}\in {\cal C}_i \times {\cal C}_j, \\
& I_{KL}\left(\boldsymbol{\theta}_{i},\boldsymbol{\mu}_{i}\right) \leq {\rho}, & \forall i \in \mathcal{N},\\ 
&  I_{KL}\left(\boldsymbol{\theta}_{ij},\boldsymbol{\mu}_{ij}\right) \leq \rho, & \forall (i,j) \in \mathcal{N}'.
\end{array}
\end{equation}


Secondly, the optimization problems (\ref{over0}) and (\ref{over1}) is closely related to the formulation of the iterative proportional fitting (IPF) procedure (see Deming and Stephan 1940, 
Fienberg 1970, 
Glasserman and Yang 2016, 
Ireland and Kullback 1968) 
which is also known as biproportional fitting or raking. In the IPF procedure, the bivariate or higher-dimension multivariate marginal information is adjusted keeping the given univariate or lower-dimension marginals respectively fixed, thereby maximizing the bivariate (or multivariate as the case may be) entropy or relative entropy which is equivalent to minimizing the KL-divergence. The optimization problem associated with the IPF method for given univariate and bivariate marginals can  as follows:
\begin{equation}\label{ren1}
\begin{array}{rll}
\displaystyle\max_{\boldsymbol{\theta}_{ij}} & -\displaystyle\sum_{(i,j) \in \mathcal{N}'} \sum_{\mathbf{c}_{ij} \in \mathcal{C}_i \times \mathcal{C}_j} \theta_{ij}(\mathbf{c}_{ij}) \;\log\left(\frac{\theta_{ij}(\mathbf{c}_{ij})}{\mu_{ij}(\mathbf{c}_{ij}}\right) \\
\textrm{subject to} & \displaystyle\sum_{c_j \in \mathcal{C}_j} \theta_{ij}(\mathbf{c}_{ij}) = \mu_i(c_i), & \forall i \in \mathcal{N}: (i,j) \in \mathcal{N}',\forall c_i \in \mathcal{C}_i, \\
& \displaystyle\sum_{c_i \in \mathcal{C}_i} \theta_{ij}(\mathbf{c}_{ij}) = \mu_j(c_j),&  \forall j \in \mathcal{N}: (i,j) \in \mathcal{N}', \forall c_j \in \mathcal{C}_j, \\
& \theta_{ij}(\mathbf{c}_{ij}) \geq 0, &  \forall (i,j) \in \mathcal{N}', \forall \mathbf{c}_{ij} \in \mathcal{C}_i \times \mathcal{C}_j.
\end{array}
\end{equation}
This problem can be reformulated in terms of the $\rho$-neighborhood formulation as follows:
\begin{equation}\label{ren2}
\begin{array}{rll}
\displaystyle\min_{\rho,\;\boldsymbol{\theta}_{ij}} & \displaystyle\sum_{(i,j) \in \mathcal{N}'} \rho_{ij} \\
\textrm{subject to} & \displaystyle\sum_{c_j \in \mathcal{C}_j} \theta_{ij}(\mathbf{c}_{ij}) = \mu_i(c_i), & \forall i \in \mathcal{N}: (i,j) \in \mathcal{N}', \forall c_i \in \mathcal{C}_i, \\
& \displaystyle\sum_{c_i \in \mathcal{C}_i} \theta_{ij}(\mathbf{c}_{ij}) = \mu_j(c_j),& \forall j \in \mathcal{N}: (i,j) \in \mathcal{N}', \forall c_j \in \mathcal{C}_j, \\
 & \displaystyle {\theta}_{ij}(\mathbf{c}_{ij}) \geq 0, & \forall (i,j) \in \mathcal{N}', \forall \mathbf{c}_{ij}\in {\cal C}_i \times {\cal C}_j, \\
&\displaystyle I_{KL}(\boldsymbol{\theta}_{ij},\boldsymbol{\mu}_{ij}) \leq \rho_{ij}, & \forall (i,j) \in \mathcal{N}'.
\end{array}
\end{equation}

\noindent Observe that the above formulation is similar to that of (\ref{over1}) except that here the objective is defined using the L$_1$-norm while in (\ref{over1}) the objective is defined using the L$_\infty$-norm.

Finally, in our model, we focus on the expert information given in terms of bivariate marginals. A related problem that has been studied is to generate distributions with given univariate marginals together with expert information specified in terms of a covariance matrix. One such method is the NORTA (NORmal To Anything) method (see Cario and Nelson 1997). 
While popular, it is known that the NORTA method  might not be able to generate joint distributions in all instances where there exists sets of univariate marginals with a feasible covariance matrix (see Ghosh and Henderson (2002)  
for counterexamples). Ghosh and Henderson (2002) 
proposed a general linear optimization formulation to construct a joint distribution with given univariate marginals and a covariance matrix if one exists or else show that the problem is infeasible. However the size of their linear program grows exponentially in the number of random variables and is hence easy to solve only in low dimensions. Using our previous result, we can extend their result as follows. Assume that, for each $(i,j) \in \mathcal{N}'$, we are given an estimate of the covariance between $i$ and $j$ represented by $\Sigma_{ij}$. Define the mean of the random variables as $E(\tilde{c}_i) = \sum_{c_i}c_i\mu_i(c_i)$. We can then generalize the formulation in (\ref{over1}) by using covariance information as follows: 
\begin{equation} \label{henderson}
\begin{array}{rll}
\displaystyle\min_{\rho,\;\boldsymbol{\theta}_{ij}} & \rho \\
\textrm{subject to} & \displaystyle\sum_{c_j \in \mathcal{C}_j} \theta_{ij}(\mathbf{c}_{ij}) = \mu_i(c_i), & \forall i \in \mathcal{N}: (i,j) \in \mathcal{N}', \forall c_i \in \mathcal{C}_i, \\
& \displaystyle\sum_{c_i \in \mathcal{C}_i} \theta_{ij}(\mathbf{c}_{ij}) = \mu_j(c_j), & \forall j \in \mathcal{N}: (i,j) \in \mathcal{N}', \forall c_j \in \mathcal{C}_j, \\
 & \displaystyle \sum_{\mathbf{c}_{ij} \in \mathcal{C}_i \times \mathcal{C}_j} {\theta}_{ij}(\mathbf{c}_{ij}) = 1, & \forall (i,j) \in \mathcal{N}',   \\ 
 & \displaystyle {\theta}_{ij}(\mathbf{c}_{ij}) \geq 0, &   \forall (i,j) \in \mathcal{N}', \forall \mathbf{c}_{ij}\in {\cal C}_i \times {\cal C}_j,   \\
&\displaystyle \left(\sum_{\mathbf{c}_{ij} \in \mathcal{C}_i \times \mathcal{C}_j} c_ic_j{\theta}_{ij}(\mathbf{c}_{ij}) - E(\tilde{c}_i)E(\tilde{c}_j)\right)-\Sigma_{ij}\leq \rho, &  \forall (i,j) \in \mathcal{N}',\\
 &\displaystyle -\left(\sum_{\mathbf{c}_{ij} \in \mathcal{C}_i \times \mathcal{C}_j} c_ic_j{\theta}_{ij}(\mathbf{c}_{ij}) - E(\tilde{c}_i)E(\tilde{c}_j)\right)+\Sigma_{ij}\leq \rho, &  \forall (i,j) \in \mathcal{N}'.
\end{array}
\end{equation}
Note that at optimality $\rho = \max_{(i,j)\in \mathcal{N}'}|E(\tilde{c}_i\tilde{c}_j)-E(\tilde{c}_i)E(\tilde{c}_j)- \Sigma_{ij}|$. If the optimal objective value in (\ref{henderson}) is $\rho = 0$, then by using a proof construction as in Theorem \ref{thm2.1}, there exists a joint distribution with the given univariate marginals and covariance matrix, else there exists no such distribution. Note that unlike the original linear program in Ghosh and Henderson (2002), 
the number of variables and constraints in the linear 
program (\ref{henderson}) is polynomial in the number of random variables. This is because we make the assumption of a tree structure on the known covariance information.



\section{Worst-Case Expected Shortfall}\label{sec3}
The distributionally robust CVaR (or expected shortfall) problem for portfolio optimization in (\ref{eq3}) is typically formulated as:
\begin{center}
$\displaystyle\min_{\mathbf{x} \in \mathcal{X},\:\beta \in \mathbb{R}} \left(\beta + \frac{1}{1-\alpha} \max_{\boldsymbol{\theta \in \Theta}}\: \mathbb{E}_{\mathbf{\theta}} [\mathbf{\tilde{c}}^T\mathbf{x} - \beta]^+ \right)$
\end{center}
In this section, we focus on solving the inner maximization problem for fixed $\mathbf{x}$ and $\beta$, and finding the worst-case distribution corresponding to it. We consider a more general expected value of functions of the form $\mathbb{E}_{\boldsymbol{\theta}} \left[\max_{k \in \mathcal{K}} \left(\mathbf{\tilde{c}}^T\mathbf{a}_k(\mathbf{x}) + b_k(\mathbf{x}) \right)\right]$ where $\mathcal{K} = \{1,2,\ldots,K\}$ and $\mathbf{a}_k(\mathbf{x})$, ${b}_k(\mathbf{x})$ are assumed to be affine functions of the decision vector $\mathbf{x} \in \mathcal{X}$. Observe that for $\mathcal{K} = \{1,2\}$, defining $\mathbf{a}_1(\mathbf{x}) = \mathbf{x}$, $b_1(\mathbf{x}) = -\beta$ and $\mathbf{a}_2(\mathbf{x}) = \mathbf{0}$, $b_2(\mathbf{x}) = 0$, the problem reduces to the inner maximization problem for the robust expected shortfall problem. For given univariate $\boldsymbol{\mu}_i$ and bivariate $\boldsymbol{\mu}_{ij}$, the set of distributions $\boldsymbol{\Theta}_{\rho}$ is defined as:

\begin{equation}\label{uset0}
\begin{array}{rl}
\boldsymbol{\Theta}_{\rho} = \Big\{\boldsymbol{\theta} : & \textrm{proj}_i(\boldsymbol{\theta}) = \boldsymbol{\mu}_i, \forall i \in \mathcal{N},\qquad \textrm{proj}_{ij}(\boldsymbol{\theta}) = \boldsymbol{\theta}_{ij}, \forall (i,j) \in \mathcal{N}',\\
& \displaystyle\sum_{\mathbf{c} \in \boldsymbol{\mathcal{C}}|c_i} \boldsymbol{\theta}(\mathbf{c}) = {\mu}_{i}(c_i),  \forall i \in \mathcal{N}, \forall c_i \in \mathcal{C}_i,\\
& \displaystyle\sum_{\mathbf{c} \in \boldsymbol{\mathcal{C}}|\mathbf{c}_{ij}} \boldsymbol{\theta}(\mathbf{c}) = {\theta}_{ij}(\mathbf{c}_{ij}), \forall (i,j) \in \mathcal{N}', \forall \mathbf{c}_{ij} \in \mathcal{C}_i \times \mathcal{C}_j,\\
& \displaystyle\sum_{\mathbf{c} \in \boldsymbol{\mathcal{C}}} \boldsymbol{\theta}(\mathbf{c}) = 1, \qquad \displaystyle \boldsymbol{\theta}(\mathbf{c}) \geq 0, \forall \mathbf{c} \in \boldsymbol{\mathcal{C}},\\
& I_{KL}(\boldsymbol{\theta}_{ij},\boldsymbol{\mu}_{ij}) \leq \rho, \forall (i,j) \in \mathcal{N}'\Big\},
\end{array}
\end{equation}
for a given $\rho \geq 0$. When no joint distribution exists corresponding to the given marginal information, $\rho$ must be at least the minimal perturbation for which the set is nonempty obtained by solving (\ref{over0}). For the case where the marginals have a consistent joint distribution, the set of distributions is nonempty for any non-negative $\rho$. The choice of $\rho$ in defining the set of distributions thus captures the confidence in the expert information. We next provide an equivalent formulation to compute the tight upper bound. The result is an extension of the linear programming problem formulation provided in Doan and Natarajan (2012) 
where we consider a $\rho$-neighborhood around the bivariate marginals thus providing for a tradeoff between conservatism in the risk measure and confidence in the expert information. The proof is provided in the appendix.

\begin{theorem}\label{thm3.1}
Consider the univariate and bivariate marginals $\left\{\boldsymbol{\mu}_i\right\}_{i \in \mathcal{N}}$ and $\left\{\boldsymbol{\mu}_{ij}\right\}_{(i,j) \in \mathcal{N}'}$ with the set ${\boldsymbol\Theta}_{\rho}$ defined in (\ref{uset0}). Let $\mathscr{V}$ denote the optimal value of the following convex program:

\begin{equation}\label{cp0}
\begin{array}{rll}
\displaystyle\max_{v^k,\;v^k_i,\;v^k_{ij},\;w_k,\;\boldsymbol{\theta},\;\boldsymbol{\theta}_{ij}} & \displaystyle\sum_{k \in \mathcal{K}} \sum_{\mathbf{c} \in \mathcal{C}} \mathbf{c}^T \mathbf{a}_k \; {v}^k(\mathbf{c}) + \sum_{k \in \mathcal{K}} b_k w_k\\
\textrm{subject to} 
& \displaystyle\sum_{\mathbf{c} \in \mathcal{C}|c_{ij}} v^k(\mathbf{c}) = v^k_{ij}(\mathbf{c}_{ij}), &  \forall (i,j) \in \mathcal{N}', \forall \mathbf{c}_{ij} \in \mathcal{C}_i \times \mathcal{C}_j, \forall k \in \mathcal{K},\\
& \displaystyle\sum_{\mathbf{c} \in \mathcal{C}|c_{i}} v^k(\mathbf{c}) = v^k_{i}(\mathbf{c}_{i}),&  \forall i \in \mathcal{N}, \forall c_i \in \mathcal{C}_i, \forall k \in \mathcal{K},\\
& \displaystyle\sum_{\mathbf{c} \in \mathcal{C}} v^k(\mathbf{c}) = w_k,&  \forall k \in \mathcal{K}, \\
 & \displaystyle\sum_{k \in \mathcal{K}} w_k = 1, & \\
%
 & \displaystyle\sum_{k \in \mathcal{K}} v^k_i(c_i) = \mu_i(c_i), &  \forall i \in \mathcal{N}, \forall c_i \in \mathcal{C}_i, \\
& \displaystyle\sum_{k \in \mathcal{K}} v^k_{ij}(\mathbf{c}_{ij}) = \theta_{ij}(\mathbf{c}_{ij}), &  \forall~(i,j) \in \mathcal{N}', \forall \mathbf{c}_{ij} \in \mathcal{C}_i \times \mathcal{C}_j,\\
 &  \displaystyle\sum_{k \in \mathcal{K}} v^k(\mathbf{c}) = \boldsymbol{\theta}(\mathbf{c}), & \forall \mathbf{c} \in \boldsymbol{\mathcal{C}}, \\ 
&  I_{KL}(\boldsymbol{\theta}_{ij},\boldsymbol{\mu}_{ij}) \leq \rho,& \forall (i,j) \in \mathcal{N}',\\
&v^k(\mathbf{c}) \geq 0, & \forall \mathbf{c} \in \mathcal{C},\forall k \in \mathcal{K}\\
& w_k \geq 0, & \forall k \in \mathcal{K}.
\end{array}
\end{equation}
Then $\mathscr{V}$ coincides with the Fr\'{e}chet bound $\mathscr{M} = \max_{\boldsymbol{\theta} \in \boldsymbol{\Theta}_{\rho}} \mathbb{E}_{\boldsymbol{\theta}} \left[\max_{k \in \mathcal{K}} \left(\mathbf{\tilde{c}}^T\mathbf{a}_k + b_k\right)\right]$.
\end{theorem}

\subsection{Tree Structure}\label{ssec3.1}
In Theorem \ref{thm3.1}, we considered the general set of distributions in (\ref{uset0}) with no additional conditions on the structure of the bivariate marginal information. Building on  Theorem \ref{thm2.1}, under the assumption of a tree structure over the set of indices $(\mathcal{N},\mathcal{N}')$, the set of distributions $\boldsymbol{\Theta}_{\rho}$ can be defined in terms of univariate and bivariate marginals as:
\begin{equation}\label{uset1}
\begin{array}{rl}
\boldsymbol{\Theta}_{\rho} = \Big\{\boldsymbol{\theta} : & \textrm{proj}_i(\boldsymbol{\theta}) = \boldsymbol{\mu}_i, \forall i \in \mathcal{N},\qquad \textrm{proj}_{ij}(\boldsymbol{\theta}) = \boldsymbol{\theta}_{ij}, \forall (i,j) \in \mathcal{N}',\\
& \displaystyle\sum_{c_j \in \mathcal{C}_j} \theta_{ij}(\mathbf{c}_{ij}) = \mu_i(c_i), \forall i \in \mathcal{N}: (i,j) \in \mathcal{N}', \forall c_i \in \mathcal{C}_i, \\
& \displaystyle\sum_{c_i \in \mathcal{C}_i} \theta_{ij}(\mathbf{c}_{ij}) = \mu_j(c_j), \forall j \in \mathcal{N}: (i,j) \in \mathcal{N}', \forall c_j \in \mathcal{C}_j, \\
& I_{KL}(\boldsymbol{\theta}_{ij},\boldsymbol{\mu}_{ij}) \leq \rho,  \forall (i,j) \in \mathcal{N}'\Big\}.
\end{array}
\end{equation}
The formulation for the worst-case bound is provided next and the proof is provided in the appendix.

\begin{theorem}\label{thm3.2}
Consider the univariate and bivariate marginals $\left\{\boldsymbol{\mu}_i\right\}_{i \in \mathcal{N}}$ and $\left\{\boldsymbol{\mu}_{ij}\right\}_{(i,j) \in \mathcal{N}'}$ such that $(\mathcal{N},\mathcal{N}')$  has a tree structure with the set $\boldsymbol{\Theta}_{\rho}$ defined in (\ref{uset1}). Let $\mathscr{V}$ denote the optimal value of the following primal convex program

\begin{equation}\label{cp1}
\begin{array}{rll}
\displaystyle\max_{v^k_i,\;v^k_{ij},\;w_k,\;\boldsymbol{\theta}_{ij}} & \displaystyle\sum_{k \in \mathcal{K}} \sum_{i \in \mathcal{N}} \sum_{{c}_{i} \in \mathcal{C}_i} {c}_{i} {a}^k_{i} \; {v}^k_{i}({c}_{i}) + \sum_{k \in \mathcal{K}} b_k w_k\\
\textrm{subject to} & \displaystyle\sum_{k \in \mathcal{K}} v^k_{ij}(\mathbf{c}_{ij}) = {\theta}_{ij}(\mathbf{c}_{ij}), &  \forall (i,j) \in \mathcal{N}', \forall \mathbf{c}_{ij} \in \mathcal{C}_i \times \mathcal{C}_j, \\
& \displaystyle\sum_{k \in \mathcal{K}} v^k_i(c_i) = \mu_i(c_i),& \forall i \in \mathcal{N}, \forall c_i \in \mathcal{C}_i, \\
& \displaystyle\sum_{\mathbf{c}_{ij} \in \mathcal{C}_{ij}} v^k_{ij}(\mathbf{c}_{ij}) = w_k,& \forall (i,j) \in \mathcal{N}',\forall k \in \mathcal{K},\\
& \displaystyle\sum_{{c}_{i} \in \mathcal{C}} v^k_{i}({c}_{i}) = w_k,& \forall i \in \mathcal{N}, \forall k \in \mathcal{K}, \\
& \displaystyle\sum_{k \in \mathcal{K}} w_k = 1, &\\
& \displaystyle\sum_{c_i \in \mathcal{C}_i} v^k_{ij}(\mathbf{c}_{ij}) = v^k_j(c_j), & \forall j \in \mathcal{N}: (i,j) \in \mathcal{N}',\forall k \in \mathcal{K}, \forall c_j \in \mathcal{C}_j, \\
& \displaystyle\sum_{c_j \in \mathcal{C}_j} v^k_{ij}(\mathbf{c}_{ij}) = v^k_i(c_i), &  \forall i \in \mathcal{N}: (i,j) \in \mathcal{N}',\forall k \in \mathcal{K}, \forall c_i \in \mathcal{C}_i, \\
 \end{array}
\end{equation}

\begin{equation*}
\begin{array}{rll}
\hspace{2cm}.& \hspace{-1.5cm} I_{KL}(\boldsymbol{\theta}_{ij},\boldsymbol{\mu}_{ij}) \leq \rho, &  \hspace{2.75cm} \forall(i,j) \in \mathcal{N}',\\
& \hspace{-1.5cm} v^k_{ij}(\mathbf{c}_{ij}) \geq 0, & \hspace{2.75cm} \forall (i,j) \in \mathcal{N}', \forall \mathbf{c}_{ij} \in \mathcal{C}_i \times \mathcal{C}_j, \forall k \in \mathcal{K},\\
& \hspace{-1.5cm} w_k \geq 0, & \hspace{2.75cm} \forall k \in \mathcal{K},
\end{array}
\end{equation*} 
Then $\mathscr{V}$ coincides with the Fr\'{e}chet bound $\mathscr{M} = \max_{\boldsymbol{\theta} \in \boldsymbol{\Theta}_{\rho}} \mathbb{E}_{\boldsymbol{\theta}} \left[\max_{k \in \mathcal{K}} \left(\mathbf{\tilde{c}}^T\mathbf{a}_k + b_k\right)\right]$.
\end{theorem}

In the formulation (\ref{cp0}), the number of decision variables is $O(Km^n)$ where $m$ is the number of possible values that each of the $n$ random variables takes whereas it is only $O(Knm^2)$ in formulation (\ref{cp1}). Thus, under the tree structure formulation, the problem is polynomial time solvable. 

In the distributionally robust CVaR problem (\ref{eq3}) which has a minmax formulation, the aim is to find the optimal decision variable $\mathbf{x}$ which minimizes the worst-case expected shortfall. Since the inner maximization problem is convex programming problem, it can be reformulated as the minimization problem using a dual formulation, thereby simplifying the distributionally robust CVaR problem to minimization problem. With the univariate and bivariate marginals $\left\{\boldsymbol{\mu}_i\right\}_{i \in \mathcal{N}}$ and $\left\{\boldsymbol{\mu}_{ij}\right\}_{(i,j) \in \mathcal{N}'}$ such that $(\mathcal{N},\mathcal{N}')$ has a tree structure along with the uncertainty set $\boldsymbol\Theta_\rho$ defined in (\ref{uset1}) with $\rho > \rho^*$ where $\rho^*$ is the optimal value of the problem (\ref{over1}), the worst-case CVaR problem (\ref{eq3}) can be reformulated as follows
\begin{equation*} 
\begin{array}{rll}
\displaystyle\min_{\mathbf{x} \in \mathcal{X},\beta,\boldsymbol{\lambda}\geq \mathbf{0},\boldsymbol{\xi},\boldsymbol{\zeta},\boldsymbol{\tau},{\nu}} & \displaystyle \left[\beta + \frac{1}{1-\alpha}\left(\nu + \sum_{(i,j) \in \mathcal{N}'} \lambda_{ij} \rho + \sum_{i \in \mathcal{N}} \sum_{c_i \in \mathcal{C}_i} \xi_i(c_i) \mu_i(c_i)\right.\right. &\\ 
& \hspace{3.5cm}\left.\left.+ \displaystyle\sum_{(i,j) \in \mathcal{N}'} \sum_{\mathbf{c}_{ij} \in \mathcal{C}_i \times \mathcal{C}_j} \lambda_{ij}\: \mu_{ij}(\mathbf{c}_{ij}) 
\left(e^{{\xi_{ij}(\mathbf{c}_{ij})}/{\lambda_{ij}}} -1\right)\right)\right]&\\
\textrm{subject to } & \nu \geq \displaystyle\sum_{i \in \mathcal{N}} \tau^1_{i}+ \displaystyle\sum_{(i,j) \in \mathcal{N}'} \tau^1_{ij} - \beta,&\\
& \nu \geq \displaystyle\sum_{i \in \mathcal{N}} \tau^2_{i}+ \displaystyle\sum_{(i,j) \in \mathcal{N}'} \tau^2_{ij},&\\
 \end{array}
\end{equation*}

\begin{equation*}
\begin{array}{rll}
& \xi_i(c_i) \geq c_i x_i - \tau_i^1 + \displaystyle\sum_{j \in \mathcal{N}:(i,j) \in \mathcal{N}'} \zeta^{1j}_i(c_i) + \sum_{l \in \mathcal{N}: (l,i) \in \mathcal{N}'} \zeta^{1l}_i(c_i), ~\forall i \in \mathcal{N},\forall c_i \in \mathcal{C}_i,& \\
& \xi_i(c_i) \geq - \tau_i^2 + \displaystyle\sum_{j \in \mathcal{N}:(i,j) \in \mathcal{N}'} \zeta^{2j}_i(c_i) + \sum_{l \in \mathcal{N}: (l,i) \in \mathcal{N}'} \zeta^{2l}_i(c_i),~\forall i \in \mathcal{N},\forall c_i \in \mathcal{C}_i,  &\\
& \xi_{ij}(\mathbf{c}_{ij}) \geq -\left( \tau^1_{ij} + \zeta^{1i}_j(c_j) + \zeta^{1j}_i(c_i)\right),~\forall (i,j) \in \mathcal{N}',\forall \mathbf{c}_{ij} \in \mathcal{C}_i \times \mathcal{C}_j, &\\
& \xi_{ij}(\mathbf{c}_{ij}) \geq -\left(\tau^2_{ij} + \zeta^{2i}_j(c_j) + \zeta^{2j}_i(c_i)\right),~\forall (i,j) \in \mathcal{N}', \forall \mathbf{c}_{ij} \in \mathcal{C}_i \times \mathcal{C}_j.&
\end{array}
\end{equation*}

\section{Numerical Experiments}\label{sec4}
In this section, we present numerical results for the Fr\'{e}chet upper bound for the inner maximization problem in the distributionally robust CVaR formulation. We consider examples wherein the complete univariate and partial bivariate information is provided with $(\mathcal{N},\mathcal{N}')$ forming a tree structure. 
Consider the worst-case upper bound:
\begin{center}
$\displaystyle\max_{\boldsymbol{\theta \in \Theta}_{\rho}}\: \mathbb{E}_{\mathbf{\theta}} \Big[\sum_{i\in \mathcal{N}} {\tilde{c}}_i - \beta\Big]^+.$
\end{center}
As the parameter $\rho$ becomes large, the bound converges to the worst-case bound obtained assuming only univariate distributions are known which is equivalent to the well-known comonotonic upper bound. For consistent marginals, as $\rho$ tends to zero, the bound converges to the worst-case bound with bivariate distributions known exactly whereas for the inconsistent scenario, as $\rho$ tends to $\rho^*$ where $\rho^*$ is the minimal perturbation obtained by solving the closest consistent marginal (\ref{over1}), the bound converges to the worst-case bound with optimal bivariate distributions for the maximum entropy problem (\ref{ren1}). Thus as $\rho$ is decreased and more confidence is attached to the expert information, the bound reduces from the comonotonic upper bound to the worst-case bound assuming the exact bivariate marginals known or that obtained from maximum entropy in the set $\mathcal{N}'$. In our computations, the bounds are estimated with the KNITRO solver that is accessed through the AMPL modeling language.


For the numerical examples we fix $\mathcal{N} = \{1,2,3,4,5\}$ with the random variable $\tilde{c}_i$ taking values in $\mathcal{C}_i = \{1,2,\ldots,10\}$ for $i \in \mathcal{N}$. Let the bivariate marginals be specified for $\mathcal{N}' = \{(1,2), (2,3), (3,4), (4,5)\}$ where the index set $(\mathcal{N},\mathcal{N}')$ form a series graph. 

We first consider an example with consistent marginal information. Corresponding to consistent univariate and bivariate marginals a joint distribution can be generated using the Chow-Liu tree distribution, thereby ensuring that the class of distributions $\boldsymbol{\Theta}_{\rho}$ defined by (\ref{uset1}) is nonempty for any $\rho \geq 0$. To compute the bound, we solve the convex optimization problem (\ref{cp1}) with $\mathcal{K} = \{1,2\}$, $a_i^1 = 1$ for $i \in \mathcal{N}$, $b_1 = -\beta$ and $a_i^2 = 0$ for $i \in \mathcal{N}$, $b_2 = 0$. Since $\sum_{i\in \mathcal{N}} {\tilde{c}}_i$ takes values in $\{5,6,\ldots,50\}$, we can vary $\beta$ in $[5,50]$. We assume that each $\tilde{c}_i$ is a discrete uniform random variable which takes values in $\mathcal{C}_i$ with probability 0.1. We consider three different kinds of dependencies modeling  expert information in the set $\mathcal{N}'$ given by, (i) very positively correlated pairs of random variables, (ii) pairwise independent random variables and (iii) very negatively correlated pairs of random variables. For positively correlated random variables, the bivariate distributions in $\mathcal{N}'$ are chosen with discrete uniform marginals and a Gaussian copula (discretized) with correlation parameter 0.69. Figure~1 represents the probability heat map of the optimal bivariate distributions for the random variables $\tilde{c}_1$ and $\tilde{c}_2$ obtained by solving the convex optimization problem (\ref{cp1}) over $\beta = 15,30,45$ corresponding to different values of the parameter $\rho = 0.00001,0.01,0.1,0.5$. We also provide the worst-case bounds in the figure (written as `Bound' in the caption of each sub-figure). Optimal distributions for the other bivariate distributions in the set $\mathcal{N}'$ are the same as observed in Figure 1 (by construction) and hence not shown here . In Figure~1, we observe that as $\rho$ decreases, the bivariate distributions are positively correlated with small changes where the bound decreases slightly. In Figure~2, we consider the case where the expert information is pairwise independent bivariate distributions (can be though of as  Gaussian copula with correlation parameter 0) in the set $\mathcal{N}'$. As the figure indicates, the worst-case bivariate distributions as $\rho$ decrease, moves from positively correlated distributions to independent distributions. The bounds decrease more in these instances as compared to Figure~1. Lastly, in Figure~3, we consider the case where the bivariate distributions in $\mathcal{N}'$ are chosen with uniform marginals and a Gaussian copula with correlation parameter --0.69. In this case, the worst-case bivariate distributions as $\rho$ decrease moves from positively correlated distributions to negatively correlated distributions. The bound reduces much more in this case since the expert information is very different from the worst-case bivariate distributions with comonotonic random variables. 


\begin{figure}\label{fig1}
\centering
\vspace{-1cm}
\includegraphics[scale=0.45]{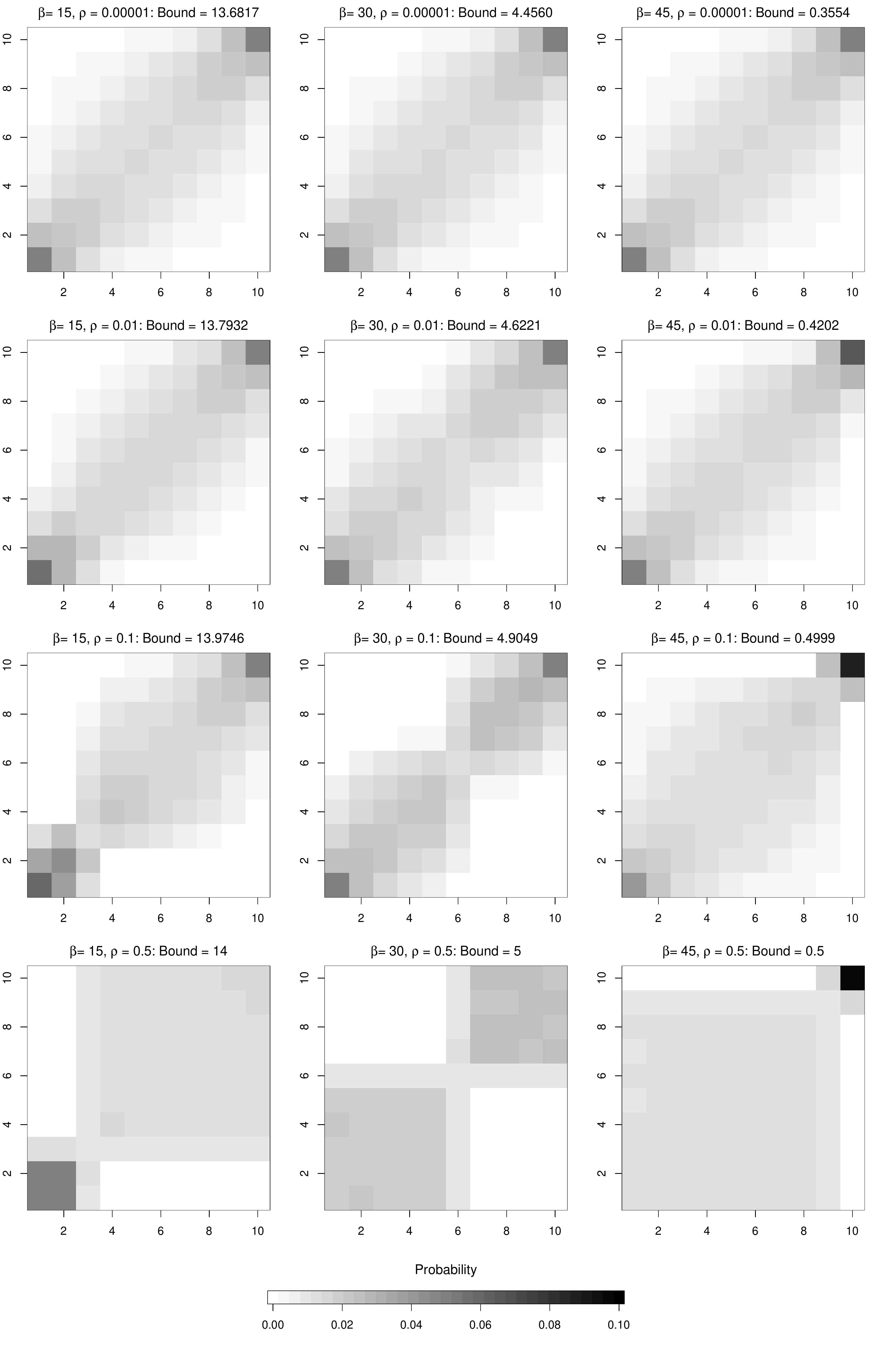}
\vspace{-0.75cm}
\hspace{1cm}\caption{Heat map for $\boldsymbol{\theta}_{12}$ for discrete uniform random variables with correlation coefficient 0.69}
\end{figure}

\begin{figure}\label{fig2}
\centering
\vspace{-1cm}
\includegraphics[scale=0.45]{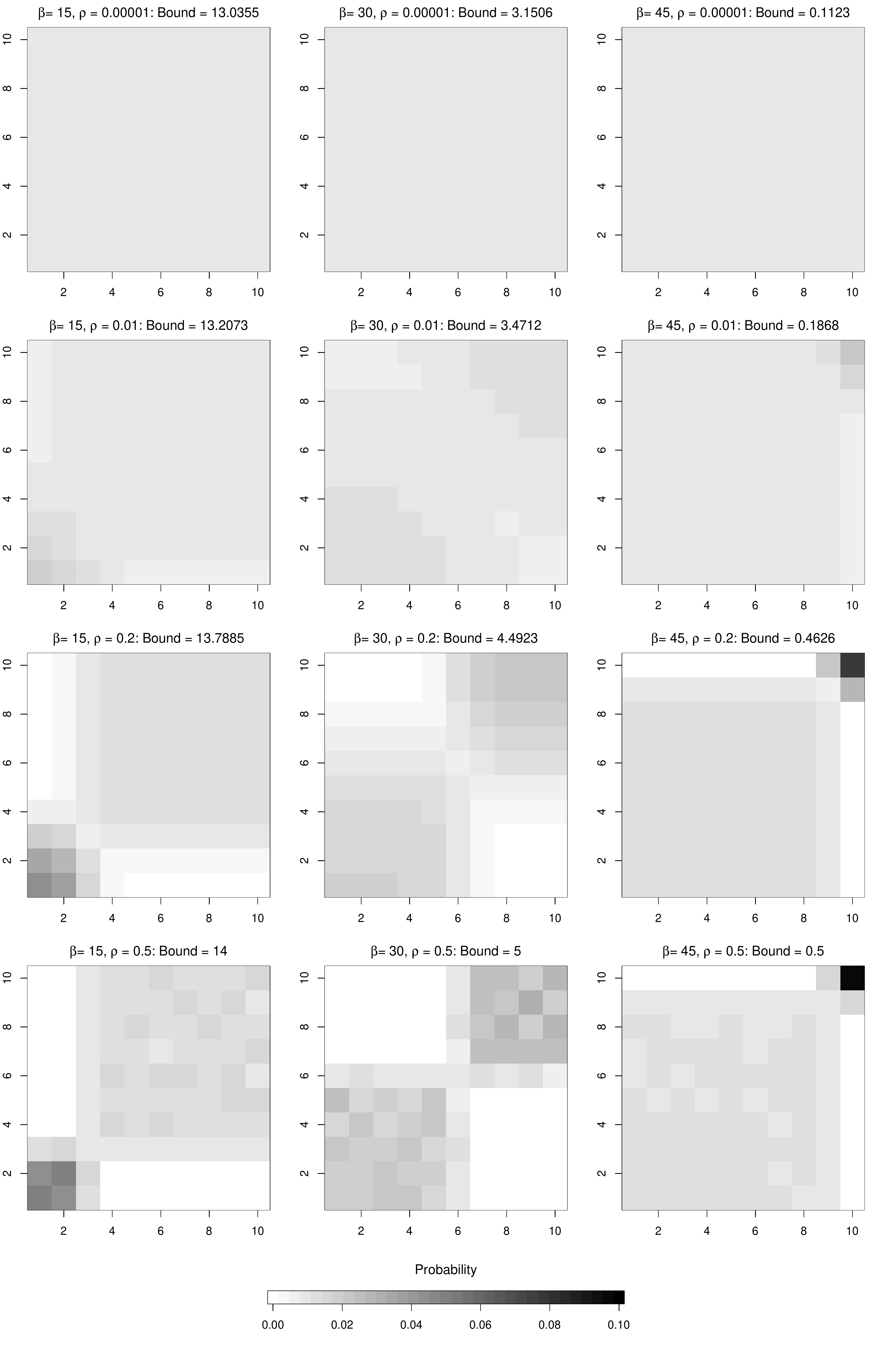}
\vspace{-0.75cm}
\hspace{1cm}\caption{Heat map for $\boldsymbol{\theta}_{12}$ for discrete uniform independent random variables}
\end{figure}

\begin{figure}\label{fig3}
\centering
\vspace{-1cm}
\includegraphics[scale=0.45]{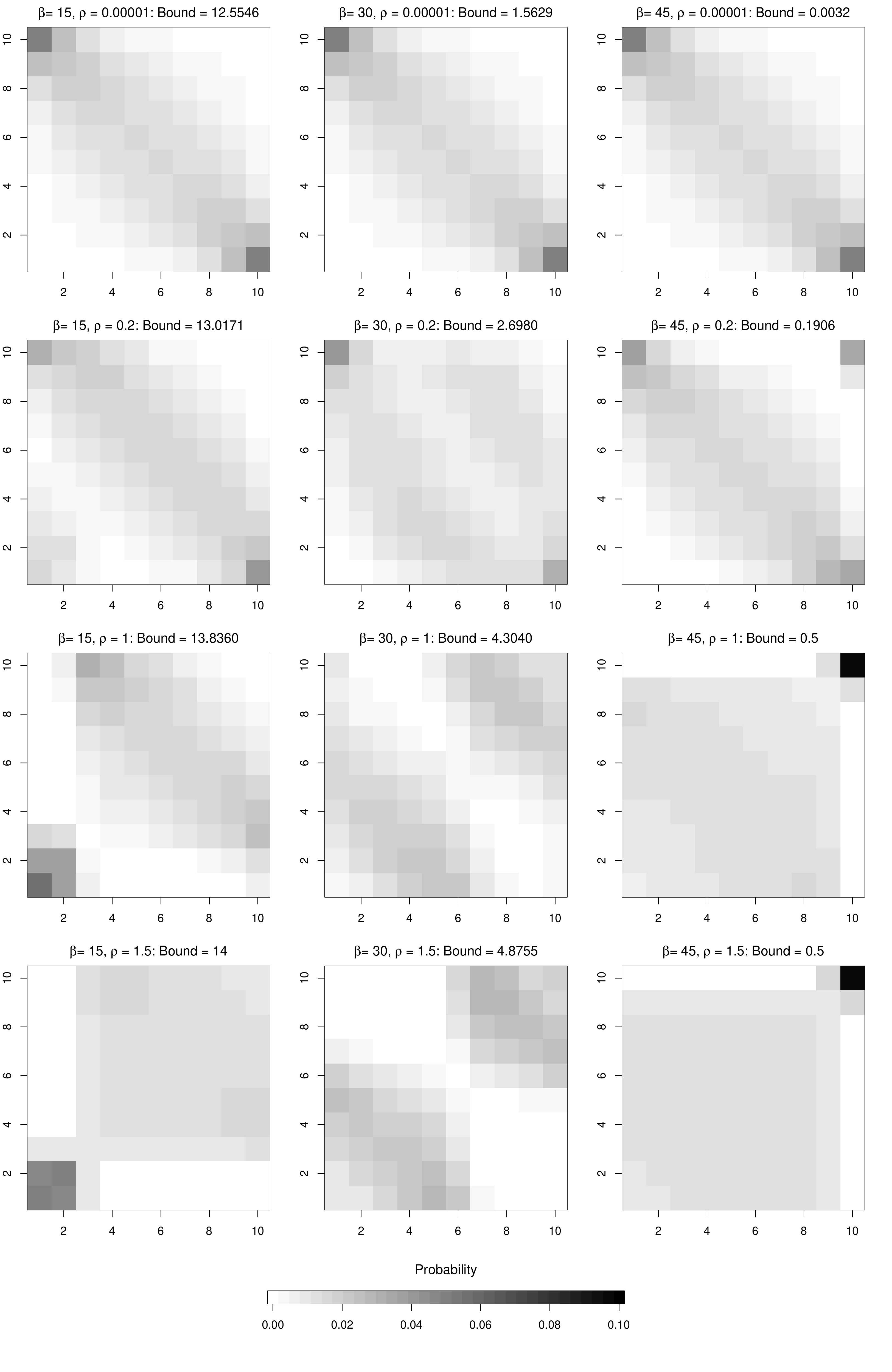}
\vspace{-0.75cm}
\hspace{1cm}\caption{Heat map for $\boldsymbol{\theta}_{12}$ for discrete uniform random variables with correlation coefficient -0.69}
\end{figure}

In Figure~4, we compare the Fr\'{e}chet bound for the perturbation parameter $\rho = 0.1$ with the bounds obtained from the bivariate distribution known exactly and from only the univariate distributions known for different $\beta$ in all the three cases. Observe that depending on the expert information (that is the bivariate distribution) known, the Fr\'{e}chet bound varies between the bounds obtained from univariate and the bivariate information. For the positively correlated scenario, the Fr\'{e}chet bound for $\rho = 0.1$ is closer to the univariate bound while for the negatively correlated case, it is closer to the bivariate bound. Thus we observe that the model is able to capture the trade-off between conservatism in the worst-case risk measure and confidence in expert information.

\begin{figure}\label{fig4}
\centering
\vspace{-1cm}
\hspace{0cm}\includegraphics[scale=0.45]{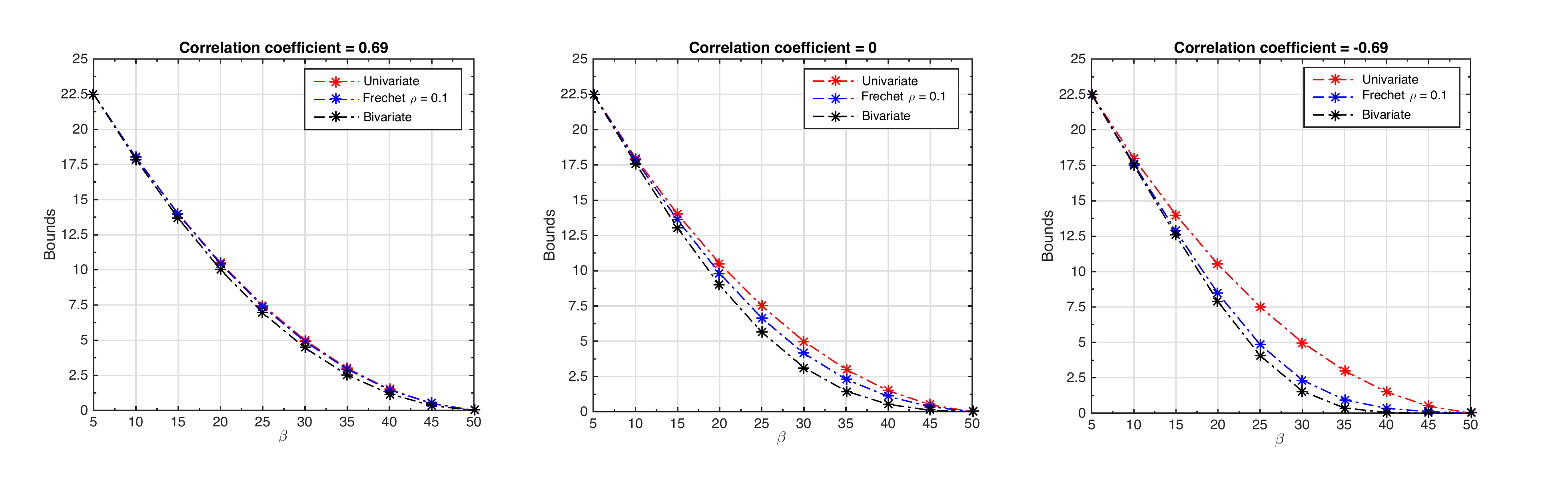}
\vspace{-0.75cm}
\hspace{1cm}\caption{Comparison of Fr\'{e}chet bound for $\rho = 0.1$ with univariate and bivariate bounds}
\end{figure}

For our next example, unlike the discrete uniform distribution above for the univariate marginals, we consider a non-uniform discrete univariate distribution for $i \in \mathcal{N}$ with the marginals given in Table~1, while generating three different bivariate distributions for $(i,j) \in \mathcal{N}'$ using the discrete Gaussian copula as in the preceding example, that is with correlation coefficients $0.69, 0$ and $-0.69$. 

\begin{table}[h]
\centering
\begin{tabular}{|c|c|c|c|c|c|c|c|c|c|c|}
\hline
$c$ & 1 & 2 & 3 & 4 & 5 & 6 & 7 & 8 & 9 & 10 \\
\hline
$\mu_i(c)$ & 0.025 & 0.050 & 0.075 & 0.15 & 0.20 & 0.20	& 0.15 & 0.075 & 0.050 & 0.025\\
\hline
\end{tabular}
\caption{Univariate marginals $\boldsymbol{\mu}_i$}\label{table1}
\end{table}
\noindent Clearly the marginals are inconsistent in all the three cases (since the discretized Gaussian copulas has discrete uniform marginals). Thus we solve the closest consistency problem (\ref{over1}) and consider the class of distribution $\boldsymbol{\Theta}_{\rho}$ with $\rho \geq \rho^*$ where $\rho^*$ is the minimal perturbation obtained. For the cases with correlation coefficient $0.69, 0$ and $-0.69$, the value of $\rho^*$ as obtained by solving (\ref{over1}) are $0.342356, 0.434234$ and $0.342356$ respectively. Moreover, the optimal bivariate distribution $\boldsymbol{\theta}_{ij}$ consistent with the given univariate as obtained by solving the relative maximum entropy problem (\ref{ren1}) are feasible for $\boldsymbol{\Theta}_{\rho}$ with $\rho \geq \rho^*$. For $\beta = 5,10,\ldots,45,50$ and different values of $\rho > \rho^*$, the Fr\'{e}chet bounds were calculated using (\ref{cp1}) which is always more than the bound calculated with given univariate and the bivariate calculated from maximum entropy. As $\rho$ increases, the Fr\'{e}chet bound converges to the univariate bound whereas as $\rho$ decreases to $\rho^*$, the Fr\'{e}chet bound converges to the maximum entropy bound. In Figure~5, we present the heat map of the optimal bivariate distributions for the random variables $\tilde{c}_1$ and $\tilde{c}_2$ obtained by solving (\ref{cp1}) over $\beta = 15,30,45$ corresponding to different values of the parameter $\rho = 0.34236,0.4,0.75,1.1$ for the case with correlation coefficient $-0.69$ while in Figure~6 we compare the Fr\'{e}chet bound for $\rho = 0.5$ with the bound obtained with maximum entropy bivariate distribution and that obtained with univariate marginals only known for all three cases. Unlike the uniform univariate marginals case in the first example, the probabilities for the optimal bivariate distributions are  concentrated towards the center; see Table ~1. Comparing Figure~5 with Figure~3 (same bivariate information, but different univariate information), we observe that although the concentration of mass is affected by the marginal univariate information provided; as $\rho$ decreases we move from a more positively correlated structure to a more negatively correlated structure in both examples. 

\begin{figure}\label{fig5}
\centering
\vspace{-1cm}
\includegraphics[scale=0.48]{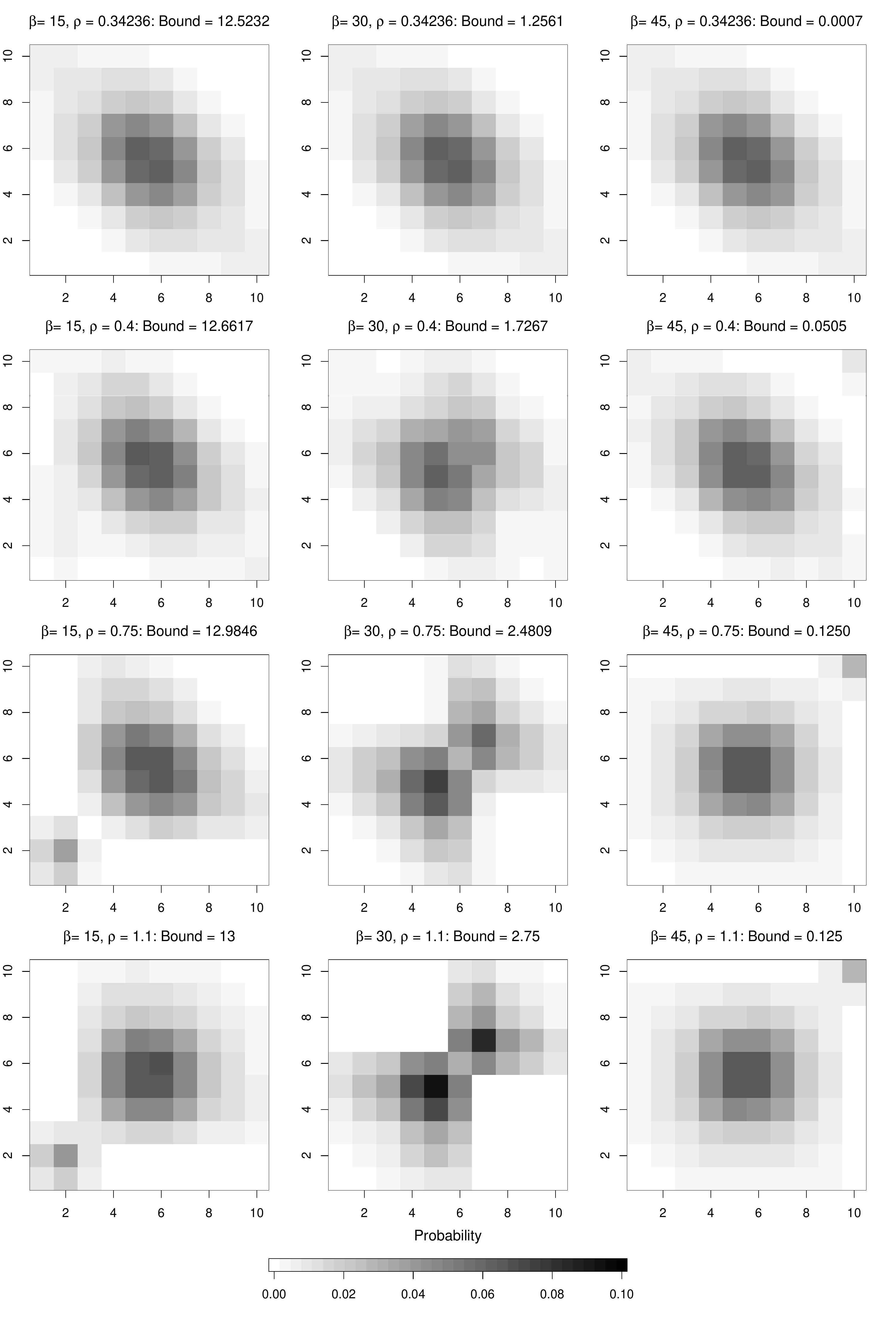}
\vspace{-0.75cm}
\hspace{1cm}\caption{Heat map for $\boldsymbol{\theta}_{12}$ for discrete random variables with correlation coefficient -0.69}
\end{figure}

\begin{figure}\label{fig6}
\centering
\vspace{-1cm}
\includegraphics[scale=0.45]{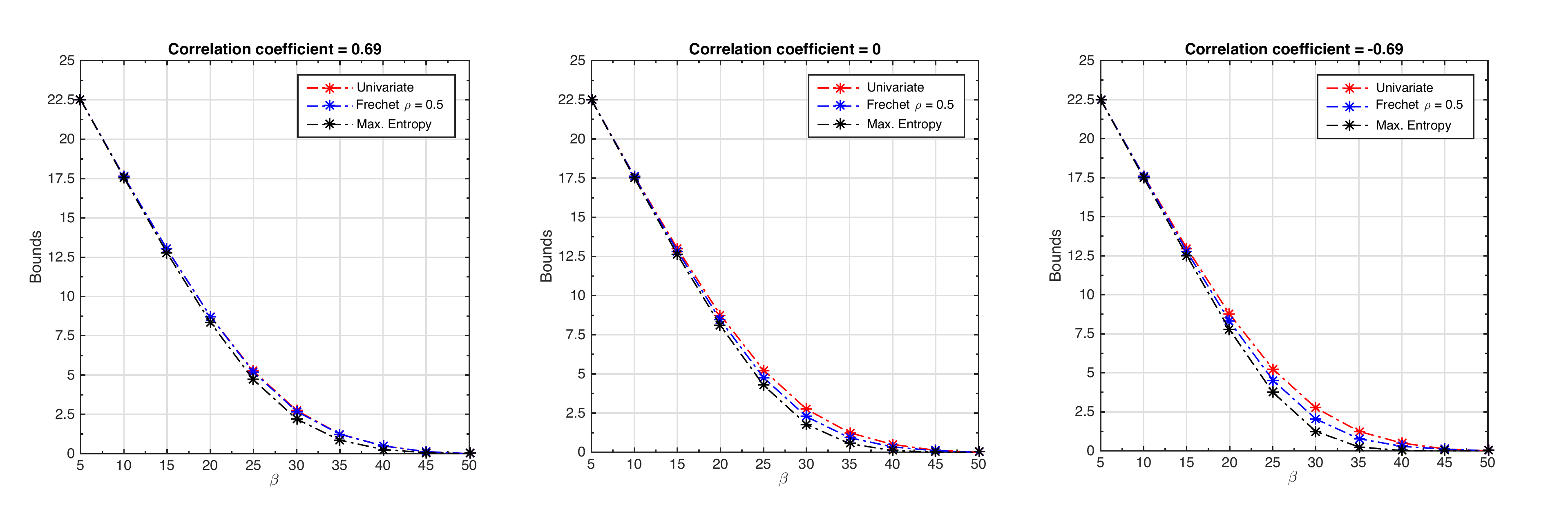}
\vspace{-0.5cm}
\hspace{1cm}\caption{Comparison of Fr\'{e}chet bound for $\rho = 0.5$ with univariate and max. entropy bounds}
\end{figure}

\section{Conclusions}\label{sec5}
In this paper, we have developed a Fr\'{e}chet upper bound for the distributionally robust CVaR problem under univariate and bivariate marginal information. By allowing for expert information in terms of bivariate marginals, we model the trade-off between conservatism in the worst-case risk measure and confidence in the expert information. Importantly, we show that as long as the bivariate marginals form a tree structure, the problem is efficiently solvable as a convex optimization problem. We end by discussing some possible directions for future research. 

In our model, we considered a class of distributions in the $\rho$-neighborhood expressed in terms of KL-divergence. Under the  KL-divergence measure, the new bivariate marginal has non-zero probability only on support points where the original bivariate marginal has non-zero probability. 
It will be interesting to extend our formulations to consider other $\phi$-divergence measures which allows one to consider bivariate marginals in a neighborhood where non-zero probabilities might be assigned to support points with zero probabilities in the expert information (see Ben-Tal et. al. 2013). Secondly, while our study has been primarily restricted to discrete distributions, it would be interesting to study how well such a method would do for continuous distributions using discretization methods. Finally, in our results, the tree structure plays a pivotal role in simplifying the formulation in comparison to the general graph structure. A natural question is whether for arbitrary bivariate information, there are methods to find an optimal tree structure for the worst-case CVaR problem.


\section*{Appendix}
\subsection*{Proof of Theorem \ref{thm2.1}}
\noindent\begin{proof}
We show the equivalence of formulations (\ref{over0}) and (\ref{over1}) when the graph associated with the univariate and bivariate marginal index sets $(\mathcal{N},\mathcal{N}')$ forms a tree structure. Given a feasible solution to (\ref{over0}) denoted by $(\rho, \boldsymbol{\theta}, \boldsymbol{\theta}_{ij})$, we consider the corresponding solution $(\rho, \boldsymbol{\theta}_{ij})$ to (\ref{over1}) with the same objective value. This solution is feasible to (\ref{over1}) since:

\begin{center}
$\displaystyle\sum_{c_j \in \mathcal{C}_j} {\theta}_{ij}(\mathbf{c}_{ij}) = \sum_{c_j \in \mathcal{C}_j} \sum_{\mathbf{c} \in \boldsymbol{\mathcal{C}}|\mathbf{c}_{ij}} \boldsymbol{\theta}(\mathbf{c}) = \sum_{\mathbf{c}\in \boldsymbol{\mathcal{C}}|c_i} \boldsymbol{\theta}(\mathbf{c})= {\mu}_i(c_i).$

$\displaystyle\sum_{c_i \in \mathcal{C}_i} {\theta}_{ij}(\mathbf{c}_{ij}) = \sum_{c_i \in \mathcal{C}_i} \sum_{\mathbf{c} \in \boldsymbol{\mathcal{C}}|\mathbf{c}_{ij}} \boldsymbol{\theta}(\mathbf{c}) = \sum_{\mathbf{c}\in \boldsymbol{\mathcal{C}}|c_j} \boldsymbol{\theta}(\mathbf{c})= {\mu}_j(c_j).$
\end{center}

\noindent By the non-negativity of $\boldsymbol{\theta}(\mathbf{c})$, the non-negativity condition of ${\theta}_{ij}(\mathbf{c}_{ij})$ holds in (\ref{over1}).

To show the converse, observe that given a feasible solution to (\ref{over1}), we can construct a joint distribution by using a conditionally independent distribution (Chow-Liu (1968) tree). For the tree structured index set $(\mathcal{N},\mathcal{N}')$, the joint distribution is defined as

\begin{center}
$\boldsymbol{\theta}(\mathbf{c}) = \displaystyle\prod_{i \in \mathcal{N}} {\mu}_i(c_i)\prod_{(i,j) \in \mathcal{N}'}\frac{{\theta}_{ij}(\mathbf{c}_{ij})}{{\mu}_i(c_i) {\mu}_j(c_j)} $
\end{center}

\noindent satisfying the univariate and bivariate marginal conditions

\begin{center}
$\displaystyle\sum_{\mathbf{c} \in \boldsymbol{\mathcal{C}}| \mathbf{c}_{ij}} \boldsymbol{\theta}(\mathbf{c}) = {\theta}_{ij}(\mathbf{c}_{ij})$ \qquad and \qquad $\displaystyle\sum_{\mathbf{c} \in \boldsymbol{\mathcal{C}}|c_i} {\boldsymbol{\theta}(\mathbf{c})} = {\mu}_{i}(c_i)$,
\end{center}

\noindent thereby satisfying the conditions of (\ref{over0}). Hence, under the tree structure (\ref{over0}) reduces to (\ref{over1}).\hfill
\end{proof}

\subsection*{Proof of Theorem \ref{thm3.1}}
\noindent\begin{proof}
Define $\psi(\mathbf{\tilde{c}}) = \max_{k \in \mathcal{K}} \left(\mathbf{\tilde{c}}^T\mathbf{a}_k + b_k\right)$. For any joint distribution $\boldsymbol{\theta} \in \boldsymbol{\Theta}_{\rho}$, we have:
\begin{center}
$\begin{array}{rl}
\mathbb{E}_{\boldsymbol{\theta}}(\psi(\mathbf{\tilde{c}})) = & \mathbb{E}_{\boldsymbol{\theta}} \Big[\max_{k \in \mathcal{K}} \left(\mathbf{\tilde{c}}^T\mathbf{a}_k + b_k\right)\Big]\\
= & \displaystyle\sum_{k \in \mathcal{K}} \mathbb{E}_{\boldsymbol{\theta}} \Big(\displaystyle \mathbf{\tilde{c}}^T\mathbf{a}_k + b_k\;\vert\; k{\textrm{-th term is maximum}}\Big)\; \mathbb{P}_{\boldsymbol{\theta}}(k{\textrm{-th term is maximum}})\\
= & \displaystyle\sum_{k \in \mathcal{K}} \sum_{\mathbf{c} \in \mathcal{C}} \mathbf{c}^T \mathbf{a}_k \; \mathbb{P}_{\boldsymbol{\theta}}(\mathbf{\tilde{c}} = \mathbf{c},\;k{\textrm{-th term is maximum}})\\
& \hspace{6cm} + \displaystyle\sum_{k \in \mathcal{K}} b_k\; \mathbb{P}_{\boldsymbol{\theta}}(k{\textrm{-th term is maximum}})\\
= & \displaystyle\sum_{k \in \mathcal{K}} \sum_{\mathbf{c} \in \mathcal{C}} \mathbf{c}^T \mathbf{a}_k \; {v}_k(\mathbf{c}) + \sum_{k \in \mathcal{K}} b_k w_k\\
\end{array}$
\end{center}

\noindent where the decision variables $v^k(\mathbf{c})$ and $w_k$ denote
\begin{center}
$v^k(\mathbf{c}) = \mathbb{P}_{\boldsymbol{\theta}}(\mathbf{\tilde{c}} = \mathbf{c},\;k{\textrm{-th term is maximum}})$ \qquad and \qquad $w_k = \mathbb{P}_{\boldsymbol{\theta}}(k{\textrm{-th term is maximum}})$.
\end{center}

\noindent Since the decision variables $v^k(\mathbf{c})$ and $w_k$ are probability measures, they are non-negative. By the definition of probability, observe that

\begin{center}
\hspace{-2.5cm}$\displaystyle\sum_{k \in \mathcal{K}} v^k(\mathbf{c}) = \sum_{k \in \mathcal{K}} \mathbb{P}_{\boldsymbol{\theta}} (\mathbf{\tilde{c}} = \mathbf{c},\;k{\textrm{-th term is maximum}}) = \mathbb{P}_{\boldsymbol{\theta}} (\mathbf{\tilde{c}} = \mathbf{c}) = \boldsymbol{\theta}(\mathbf{c})$,

$ \displaystyle\sum_{\mathbf{c} \in \mathcal{C}} v^k(\mathbf{c}) = \sum_{\mathbf{c} \in \mathcal{C}} \mathbb{P}_{\boldsymbol{\theta}} (\mathbf{\tilde{c}} = \mathbf{c},\;k{\textrm{-th term is maximum}}) = \mathbb{P}_{\boldsymbol{\theta}} (k{\textrm{-th term is maximum}}) = w_k$,

\hspace{-6.75cm}$\displaystyle\sum_{k \in \mathcal{K}} w_k = \sum_{k \in \mathcal{K}} \mathbb{P}_{\boldsymbol{\theta}} (k{\textrm{-th term is maximum}}) =1$.
\end{center}

\noindent Further we introduce the following decision variables

\begin{center}
  $v^k_{ij}(\mathbf{c}_{ij}) = \mathbb{P}_{\boldsymbol{\theta}}(\mathbf{\tilde{c}}_{ij} = \mathbf{c}_{ij},\;k{\textrm{-th term is maximum}})$ ~ and ~ $v^k_i(c_i) = \mathbb{P}_{\boldsymbol{\theta}}({\tilde{c}}_{i} = {c}_{i},\;k{\textrm{-th term is maximum}})$. 
\end{center}

\noindent The variables $v^k(\mathbf{c})$, $v^k_{ij}(\mathbf{c}_{ij})$ and $v^k_i(c_i)$ are related as follows:
\begin{center}
\hspace{-2.5cm}$\displaystyle\sum_{\mathbf{c} \in \mathcal{C}|\mathbf{c}_{ij}} v^k(\mathbf{c}) = \sum_{\mathbf{c} \in \mathcal{C}|\mathbf{c}_{ij}} \mathbb{P}_{\boldsymbol{\theta}} (\mathbf{\tilde{c}} = \mathbf{c},\;k{\textrm{-th term is maximum}})$

\hspace{3.75cm}$ = \mathbb{P}_{\boldsymbol{\theta}} (\mathbf{\tilde{c}}_{ij} = \mathbf{c}_{ij},\;k{\textrm{-th term is maximum}}) = v^k_{ij}(\mathbf{c}_{ij}),$

\hspace{-2.5cm}$\displaystyle\sum_{\mathbf{c} \in \mathcal{C}|c_{i}} v^k(\mathbf{c}) = \sum_{\mathbf{c} \in \mathcal{C}|c_{i}} \mathbb{P}_{\boldsymbol{\theta}} (\mathbf{\tilde{c}}  = \mathbf{c},\;k{\textrm{-th term is maximum}}) $

\hspace{3.25cm}$= \mathbb{P}_{\boldsymbol{\theta}} ({\tilde{c}}_{i} = {c}_{i},\;k{\textrm{-th term is maximum}}) = v^k_{i}({c}_{i}).$
\end{center}

\noindent Moreover, $v^k_{ij}(\mathbf{c}_{ij})$ and $v^k_i(c_i)$ are related to $\theta_{ij}(\mathbf{c}_{ij})$ and $\mu_i(c_i)$ respectively as

\begin{center}
$\displaystyle\sum_{k \in \mathcal{K}} v^k_{ij}(\mathbf{c}_{ij}) =  \displaystyle\sum_{k \in \mathcal{K}} \mathbb{P}_{\boldsymbol{\theta}}(\mathbf{\tilde{c}}_{ij} = \mathbf{c}_{ij},\; k{\textrm{-th term is maximum}}) = \mathbb{P}_{\boldsymbol{\theta}}(\mathbf{\tilde{c}}_{ij} = \mathbf{c}_{ij}) = \theta_{ij}(\mathbf{c}_{ij})$

$\displaystyle\sum_{k \in \mathcal{K}} v^k_i(c_i) =  \displaystyle\sum_{k \in \mathcal{K}} \mathbb{P}_{\boldsymbol{\theta}}(\tilde{c}_i = c_i,\; k{\textrm{-th term is maximum}}) = \mathbb{P}_{\boldsymbol{\theta}}(\tilde{c}_i = c_i) = \mu_i(c_i)$
\end{center}

\noindent For the given univariate and bivariate marginals $\boldsymbol{\mu}_i$ and $\boldsymbol{\mu}_{ij}$, $\boldsymbol{\theta}_{ij}$ satisfy the following $\rho$-neighborhood condition 

\begin{center}
$I_{KL}(\boldsymbol{\theta}_{ij}, \boldsymbol{\mu}_{ij}) \leq \rho,~\quad \forall~(i,j) \in \mathcal{N}'$
\end{center}

\noindent such that it is consistent with $\boldsymbol{\mu}_i$. Thus for any $\boldsymbol{\theta} \in \boldsymbol{\Theta}_{\rho}$ along with the decision variables $\boldsymbol{v}_k,\;\boldsymbol{v}^k_i,\; \boldsymbol{v}^k_{ij},\; w_k$ satisfying the constraints, 

\begin{center}
$\displaystyle\max_{\boldsymbol{v}_k,\boldsymbol{v}^k_{ij},\boldsymbol{v}^k_i,w_k}\;\sum_{k \in \mathcal{K}} \sum_{\mathbf{c} \in \mathcal{C}} \mathbf{c}^T \mathbf{a}_k \; {v}_k(\mathbf{c}) + \sum_{k \in \mathcal{K}} b_k w_k \geq \mathbb{E}_{\boldsymbol{\theta}}(\psi({\mathbf{\tilde{c}}}))$,
\end{center}

\noindent which implies 

\begin{center}
$\mathscr{V} =  \displaystyle\max_{\boldsymbol{v}_k,\boldsymbol{v}^k_{ij},\boldsymbol{v}^k_i,w_k,\boldsymbol{\theta}}\;\sum_{k \in \mathcal{K}} \sum_{\mathbf{c} \in \mathcal{C}} \mathbf{c}^T \mathbf{a}_k \; {v}_k(\mathbf{c}) + \sum_{k \in \mathcal{K}} b_k w_k \geq \displaystyle\max_{\boldsymbol{\theta}\in \boldsymbol{\Theta}} \mathbb{E}_{\boldsymbol{\theta}}(\psi(\mathbf{\tilde{c}}) =  \mathscr{M}$.
\end{center}

To prove the result, we need to show that the bound is tight, that is $\mathscr{V} = \mathscr{M}$. Consider the optimal decision variables $\boldsymbol{v}_k^*,\;\boldsymbol{v}^{k*}_{ij},\;\boldsymbol{v}^{k*}_i,\;w^*_k,\;\boldsymbol{\theta}^*$ and $\boldsymbol{\theta}_{ij}^*$ of the convex programming problem (\ref{cp0}). Observe that $w^*_k$ is a probability measure. Next for a fixed $k \in \mathcal{K}$, we construct distributions $\boldsymbol{\theta}^{k*}$ and $\boldsymbol{\theta}^{k*}_{ij}$ as follows:

\begin{itemize}
\item[(a)] Choose the $k$-th term $\mathbf{\tilde{c}}^T\mathbf{a}_k + b_k$ with probability $w^*_k$.

\item[(b)]  For a fixed $k \in \mathcal{K}$, define $\boldsymbol{\theta}^{k*}(\mathbf{c}) = v^*_k(\mathbf{c})/w^*_k$. In addition, for each $(i,j) \in \mathcal{N}'$, define $\theta^{k*}_{ij}(\mathbf{c}_{ij}) = v^{k*}_{ij}(\mathbf{c}_{ij})/w^*_k$ and for $i \in \mathcal{N}$, $\theta^{k*}_{i}({c}_{i}) = v^{k*}_{i}({c}_{i})/w^*_k$. Note that if $w^*_k = 0$, we simply drop that index. It is easy to observe that 

\begin{center}
$\displaystyle\sum_{\mathbf{c} \in \mathcal{C}|\mathbf{c}_{ij}} \boldsymbol{\theta}^{k*}(\mathbf{c}) = \theta^{k*}_{ij}(\mathbf{c}_{ij})$ \quad and \quad $\displaystyle\sum_{\mathbf{c} \in \mathcal{C}|{c}_{i}} \boldsymbol{\theta}^{k*}(\mathbf{c}) = \theta^{k*}_{i}({c}_{i})$.
\end{center}
\end{itemize}

\noindent Hence the following inequality holds:

\begin{center}
$\mathbb{E}_{\boldsymbol{\theta}^{k*}}\bigg[\displaystyle\max_{l \in \mathcal{K}} \Big(\mathbf{\tilde{c}}^T \mathbf{a}_l + b_l\Big)\bigg] \geq  \mathbb{E}_{\boldsymbol{\theta}^{k*}}\Big[\displaystyle\mathbf{\tilde{c}}^T \mathbf{a}_k + b_k\Big] = \displaystyle\frac{1}{w^*_k}\sum_{\mathbf{c} \in \mathcal{C}} \mathbf{{c}}^T\mathbf{a}_k v^{k*}(\mathbf{c}) + b_k$
\end{center}

\noindent where the first inequality is obtained by simply choosing the $k$-th term in the function for $\boldsymbol{\theta}^{k*}$. Since

\begin{center}
$\begin{array}{rl}
\mathbb{E}_{\boldsymbol{\theta}^*}\bigg[\displaystyle\max_{l \in \mathcal{K}} \Big(\mathbf{\tilde{c}}^T \mathbf{a}_l + b_l\Big)\bigg] = & \displaystyle\sum_{k \in \mathcal{K}} w^*_k \mathbb{E}_{\boldsymbol{\theta}^{k*}}\bigg[\displaystyle\max_{l \in \mathcal{K}} \Big( \mathbf{\tilde{c}}^T\mathbf{a}_l + b_l \Big)\bigg]\\
\geq & \displaystyle\sum_{k \in \mathcal{K}} w^*_k \Big[\frac{1}{w^*_k}\sum_{\mathbf{c} \in \mathcal{C}} \mathbf{\tilde{c}}^T \mathbf{a}_k v^{k*}(\mathbf{c}) + b_k\Big] 
=  \displaystyle\sum_{k \in \mathcal{K}} \sum_{\mathbf{c} \in \mathcal{C}} \mathbf{\tilde{c}}^T \mathbf{a}_k v^{k*}(\mathbf{c}) + \sum_{k \in \mathcal{K}} b_k w^*_k.
\end{array}$
\end{center}

\noindent Therefore, $\mathscr{M} = \displaystyle\max_{\boldsymbol{\theta} \in \boldsymbol{\Theta}_{\rho}} \:\mathbb{E}_{\boldsymbol{\theta}} \Big[\max_{k \in \mathcal{K}} \big(\mathbf{\tilde{c}}^T\mathbf{a}_k + b_k\big)\Big] \geq \mathbb{E}_{\boldsymbol{\theta}^*} \Big[\max_{k \in \mathcal{K}} \big(\mathbf{\tilde{c}}^T\mathbf{a}_k + b_k\big)\Big] \geq \mathscr{V}$ which together with the fact that $\mathscr{V} \geq \mathscr{M}$ ensures that $\mathscr{V} = \mathscr{M}$.\hfill
\end{proof}

\subsection*{Proof of Theorem \ref{thm3.2}}
\noindent\begin{proof}
Define $\psi(\mathbf{\tilde{c}}) = \max_{k \in \mathcal{K}} \left(\mathbf{\tilde{c}}^T\mathbf{a}_k + b_k\right)$. For any joint distribution $\boldsymbol{\theta} \in \boldsymbol{\Theta}_{\rho}$, the expected value is expressed as:
\begin{center}
$\mathbb{E}_{\boldsymbol{\theta}}(\psi(\mathbf{\tilde{c}})) 
= \displaystyle\sum_{k \in \mathcal{K}} \sum_{i \in \mathcal{N}} \sum_{{c}_{i} \in \mathcal{C}_i} {c}_{i} {a}^k_{i} \; {v}^k_{i}({c}_{i}) + \sum_{k \in \mathcal{K}} b_k w_k$
\end{center}

\noindent where the decision variable $v^k_{i}({c}_{i})$ and $w_k$ denote
\begin{center}
$v^k_{i}({c}_{i}) = \mathbb{P}_{\boldsymbol{\theta}}({\tilde{c}}_{i} = {c}_{i},\;k{\textrm{-th term is maximum}})$ \quad and \quad $w_k = \mathbb{P}_{\boldsymbol{\theta}}(k{\textrm{-th term is maximum}})$
\end{center}

\noindent Further we introduce another decision variable $v^k_{ij}(\mathbf{c}_{ij}) = \mathbb{P}_{\boldsymbol{\theta}}({\mathbf{\tilde{c}}}_{ij} = \mathbf{c}_{ij},\;k{\textrm{-th term is maximum}})$ to incorporate the consistency relation between the given univariate $\boldsymbol{\mu}_i$ and the perturbed bivariate $\boldsymbol{\theta}_{ij}$. Using the basic properties of probability measures, the constraints are easily satisfied.
Thus for any $\boldsymbol{\theta} \in \boldsymbol{\Theta}_{\rho}$ along with the decision variables $\boldsymbol{v}^k_i,\; \boldsymbol{v}^k_{ij},\; w_k$ satisfying the constraints, 

\begin{center}
$\mathscr{V} =  \displaystyle\max_{\boldsymbol{v}^k_{ij},\boldsymbol{v}^k_i,w_k,\boldsymbol{\theta}}\;\sum_{k \in \mathcal{K}} \sum_{i \in \mathcal{N}} \sum_{{c}_{i} \in \mathcal{C}_i} {c}_{i} {a}^k_{i} \; {v}^k_{i}({c}_{i}) + \sum_{k \in \mathcal{K}} b_k w_k \geq \: \displaystyle\max_{\boldsymbol{\theta}\in \boldsymbol{\Theta}}\: \mathbb{E}_{\boldsymbol{\theta}}(\psi(\mathbf{\tilde{c}})) =  \mathscr{M}$.
\end{center}

For the reverse inequality $\mathscr{M} \geq \mathscr{V}$, consider the optimal decision variables $\boldsymbol{v}^{k*}_{ij}, ~\boldsymbol{v}^{k*}_i$ and $w^*_k$ of the convex programming problem. Next for a fixed $k \in \mathcal{K}$ we construct a joint distribution $\boldsymbol{\theta}^{k*}$ as follows:

\begin{itemize}
\item[(a)] Choose the $k$-th term $\sum_{i \in \mathcal{N}} \sum_{{c}_{i} \in \mathcal{C}_i} {c}_{i} {a}^k_{i} + b_k$ with probability $w^*_k$.

\item[(b)]  For each $(i,j) \in \mathcal{N}'$, define $\theta^{k*}_{ij}(\mathbf{c}_{ij}) = v^{k*}_{ij}(\mathbf{c}_{ij})/w^*_k$ and for each $i \in \mathcal{N}$, define $\theta^{k*}_{i}({c}_{i}) = v^{k*}_{i}({c}_{i})/w^*_k$. Note that if $w^*_k = 0$, we simply drop that index. Using the consistency conditions of $v^{k*}_{ij}(\mathbf{c}_{ij})$ and $v^{k*}_i(c_i),\;v^{k*}_j(c_j)$, it is easy to observe that $\theta^{k*}_{ij}(\mathbf{c}_{ij})$ and $\theta^{k*}_i(c_i),\;\theta^{k*}_j(c_j)$ are consistent.

\item[(c)] For $k \in \mathcal{K}$, since $(\mathcal{N},\mathcal{N}')$ forms a tree structure, there exists a consistent joint distribution $\boldsymbol{\theta}^{k*}$ corresponding to $\theta^{k*}_{ij}(\mathbf{c}_{ij})$ and $\theta^{k*}_i(c_i),\;\theta^{k*}_j(c_j)$.
\end{itemize}

\noindent Now working along the lines of the proof of Theorem \ref{thm3.1}, we obtain

\begin{center}
$\mathbb{E}_{\boldsymbol{\theta}^*}\bigg[\displaystyle\max_{l \in \mathcal{K}} \Big(\mathbf{\tilde{c}}^T\mathbf{a}_l + b_l\Big)\bigg] \geq \displaystyle\sum_{k \in \mathcal{K}} \sum_{i \in \mathcal{N}} \sum_{{c}_{i} \in \mathcal{C}_i} {{c}}_{i}{a}^k_{i} v^{k*}_{i}({c}_{i}) + \sum_{k \in \mathcal{K}} b_k w^*_k.$
\end{center}

\noindent Therefore, $\mathscr{M}\geq \mathscr{V}$ which together with the fact that $\mathscr{V} \geq \mathscr{M}$ ensures that $\mathscr{V} = \mathscr{M}$.\hfill
\end{proof}

\section*{References}

\noindent Artzner, P., F. Delbaen, J-M. Eber, and D. Heath (1999): {Coherent measures of risk}, \emph{Mathematical Finance}, 9, 203-228.

\noindent Bedford, T., and R. M. Cooke (2002): {A new graphical model for dependent random variables}, \emph{Annals of Statistics}, 30, 1031-1068.

\noindent Ben-Tal, A., D. den Hertog, A. De Waegenaere, B. Melenberg, and G. Rennen (2013): {Robust solutions of optimization problems affected by uncertain probabilities}, \emph{Management Science}, 59, 341-357.

\noindent Cario, M. C., and B. L. Nelson (1997): Modeling and generating
random vectors with arbitrary marginal distributions
and correlation matrix, \emph{Technical Report}, Department
of Industrial Engineering and Management Sciences,
Northwestern University, Evanston, IL.

\noindent Chow, C. K., and C. N. Liu (1968): {Approximating discrete probability distributions with dependence trees}, \emph{IEEE Transactions on Information Theory}, 14, 462-467.

\noindent Deming, W. E., and F. F. Stephan (1940): {On a least squares adjustment of a sampled frequency table when the expected marginal totals are known}, \emph{The Annals of Mathematical Statistics}, 11, 427-444.

\noindent Doan, X. V., and K. Natarajan (2012): {On the complexity of nonoverlapping multivariate marginal bounds for probabilistic combinatorial optimization problems}, \emph{Operations Research}, 60, 138-149.

\noindent Doan, X. V., X. Li, and K. Natarajan (2015): {Robustness to dependency in portfolio optimization using overlapping marginals}, \emph{Operations Research}, 63, 1468-1488.

\noindent Embrechts, P., and G. Puccetti (2006): {Bounds for functions of multivariate risks}, \emph{Journal of Multivariate Analysis}, 97, 526-547.

\noindent Embrechts, P., G. Puccetti, L. R\"{u}schendorf, R. Wang, and A. Beleraj (2014): {An academic response to Basel 3.5}, \emph{Risks}, 2, 25-58.

\noindent Fienberg, S. E. (1970): {An iterative procedure for estimation in contingency tables}, \emph{The Annals of Mathematical Statistics}, 41, 907-917.

\noindent Ghosh, S., and S. G. Henderson (2002): {Chessboard Distributions and Random Vectors with Specified Marginals and Covariance Matrix}, \emph{Operations Research}, 50, 820-834.

\noindent Glasserman, P., and L. Yang (2016): {Bounding wrong-way risk in CVA calculation}, \emph{Mathematical Finance}, doi:10.1111/mafi.12141.

\noindent Hanasusanto, G. A., D. Kuhn, and W. Wiesemann (2016): {A comment on ``Computational complexity of stochastic programming problems''},
\emph{Mathematical Programming}, doi:10.1007/s10107-015-0958-2.

\noindent Higham, N. J. (2002): {Computing the nearest correlation matrix-a problem from finance}, \emph{IMA Journal of Numerical Analysis}, 22, 329-343.

\noindent Ireland, C. T., and S. Kullback (1968): {Contingency tables with given marginals}, \emph{Biometrika}, 55, 179-188.

\noindent Jaynes, E. T. (1957): {Information theory and statistical mechanics}, \emph{Physical Review}, 106, 620-630.

\noindent Lim, A. E. B., J. G. Shanthikumar, and G.-Y. Vahn (2011): {Conditional value-at-risk in portfolio optimization: Coherent but fragile}, \emph{Operations Research Letters}, 39, 163-171.

\noindent Pardo, L. (2006): \emph{Statistical inference based on divergence measures},  Statistics: Textbooks and Monographs, 185, Boca Raton: Chapman \& Hall/CRC FL.

\noindent Qi, H. D., and D. Sun (2010): {Correlation stress testing for Value-at-Risk: An unconstrained convex optimization approach}, \emph{Computational Optimization and Applications}, 45, 427-462.

\noindent Rockafellar, R. T., and S. Uryasev (2002): {Conditional value-at-risk for general loss distributions}, \emph{Journal of Banking \& Finance}, 26, 1443-1471.

\noindent Roughgarden, T., and M. Kearns (2013): {Marginals-to-models reducibility}, \emph{Proceedings of the 26th International Conference on Neural Information Processing Systems}, Lake Tahoe, Nevada, 1043-1051.

\noindent R\"{u}schendorf, L. (1983): {Solution of a statistical optimization problem by rearrangement method}, \emph{Metrika}, 30, 55-61.

\noindent Shannon, C. E. (1948): {A mathematical theory of communication}, \emph{Bell System Technical Journal}, 27, 379-423, 623-656.

\noindent Sklar, A. (1959): {Fonctions de r\'{e}partition \`{a} n dimensions et leurs marges}, \emph{Publications de l’Institut de Statistique de L’Universit\'{e} de Paris}, 8, 229-231.

\noindent Vorob'ev, N. N. (1962): {Consistent families of measures and their extensions}, \emph{Theory of Probability and Its Applications}, 7, 147-163.

\noindent Zhu, S., and M. Fukushima (2009): {Worst-case Conditional Value-at-Risk with application to robust portfolio management}, \emph{Operations Research}, 57, 1155-1168.

\end{document}